\documentstyle[epsf]{ioplppt}                

\begin{document}

\chardef\iacc="10
\newcommand{\rev}[1]{{{#1}}}
\newcommand{\boldm}[1]{\mbox{\protect\boldmath$#1$}}
\newcommand{\bm}[1]{\mbox{\protect\bf{#1}}}
\newcommand{\rd}{{\rm{d}}}
\newcommand{\rD}{{\rm{D}}}
\newcommand{\calS}{{S}}
\newcommand{\Phim}{{\Phi_{\rm{m}}}}
\newcommand{\Phie}{{\Phi_{\rm{e}}}}
\newcommand{\beq}{\begin{equation}}
\newcommand{\eeq}{\end{equation}}
\def\lb#1{{\protect\linebreak[#1]}}
\def\spose#1{\hbox to 0pt{#1\hss}}
\def\lta{\mathrel{\spose{\lower 3pt\hbox{$\mathchar"218$}}
     \raise 2.0pt\hbox{$\mathchar"13C$}}}
\def\gta{\mathrel{\spose{\lower 3pt\hbox{$\mathchar"218$}}
     \raise 2.0pt\hbox{$\mathchar"13E$}}}
\protect\hyphenation{elec-tro-mag-net-ic}

\title{Magnetic fields around black holes}[Magnetic fields around black
 holes]

\author{M Dov\v{c}iak\dag, V Karas\dag\ and A Lanza\ddag}

\address{\dag\ Astronomical Institute, Charles University Prague,
 V~Hole\v{s}ovi\v{c}k\'ach~2, CZ-180\,00~Praha, Czech~Republic}

\address{\ddag\ International School for Advanced Studies, Via Beirut
 2/4, I-34014~Trieste, Italy
 \medskip

 \noindent
 E-mail:
 dovciak@bb.sanet.sk;
 vladimir.karas@mff.cuni.cz;
 lanza@sissa.it
}

\begin{abstract}
Mutual interaction between electromagnetic and gravitational fields can
be treated in the framework of the general theory of relativity. This
poses a difficult problem but different approximations can be introduced
in order to grasp interesting effects and to describe mechanisms which
govern astronomical objects.

The electromagnetic field near a rotating black hole is being explored
here. By employing an analytic solution for the weak electromagnetic
field in vacuum, we plot the surfaces of constant flux and we show how
the field is dragged around the black hole by purely geometrical effects
of the strong gravitational field. We visualize the complicated
structure of magnetic lines and we also mention possible astrophysical
applications of more realistic situations involving the presence of
plasma. The entangled and twisted field lines result in reconnection
processes which accelerate the particles injected into the vicinity of the
black hole. Such acceleration mechanisms of the electromagnetic origin are
considered as a possible source of high-speed plasma jets emerging
from numerous nuclei of galaxies where black holes reside.
\end{abstract}

\pacs{04.20.q, 95.30.Sf, 97.60.L}

\bigskip

\noindent
European Journal of Physics, in press

\maketitle

\section{Introduction}
Magnetic fields play an important role in astrophysics. Near rotating
bodies, e.g.\ neutron stars and black holes, the field lines are
deformed by a rapidly moving plasma and strong gravitational fields.
Here we will illustrate purely gravitational effects by exploring
simplified vacuum solutions in which the influence of plasma is ignored
but the presence of strong gravity is taken into account.

The electromagnetic field is governed by Maxwell's equations. These are
the first-order differential equations for the electric and magnetic
intensity vectors. When expressed in the equivalent and elegant
tensorial formalism, the mutually coupled equations for the field
intensities (electric and magnetic) can be unified in terms of the
electromagnetic field tensor, comprising both the electric and the
magnetic components in a single quantity. As is well known, the unifying
approach turns out particularly useful in the framework of the theory of
relativity. Whichever formulation is preferred, one has to tackle
differential equations and, therefore, the appropriate initial and
boundary conditions must be specified in order to determine the
structure of the field completely.


Here, in this paper we deal with such a stationary state because it is
simple and has the capability of illustrating basic properties of
non-radiating fields in a clear way. Various examples have been examined
in textbooks on classical electrodynamics (Jackson 1975) where one of
the simplest illustrative cases concerns a rotating sphere immersed in
an external uniform magnetic field (for further references, see Krotkov
\etal\ 1999). According to the intuitive definition, the uniform (or
homogeneous) field is characterized by field lines which are parallel to
each other (we will soon see that a more precise definition is necessary
because the field lines are not a physical entity and their shape is
observer dependent). Quite naturally, the uniform field can be
maintained by fixed boundary conditions far away from the rotating
sphere. But the magnetic lines do reflect the presence of the body,
their shape being distorted near its surface in accordance with the
material properties (Bullard 1949; Herzenberg \& Lowes 1957). This
classical phenomenon was discussed in original works by Faraday, Lamb,
Thomson and Hertz, and it has found numerous astrophysical and
geophysical applications. Notice at this point that both the field
structure far from the body, and the rotation of the body itself are
supposed to be kept constant. Otherwise, the sphere would continuously
slow down due to dissipation of energy by Foucault currents.

Here we explore an analogous but somewhat more complicated situation: a
rotating black hole instead of the classical sphere. This means that we
either need to employ a solution of the coupled Einstein-Maxwell
equations for the gravitational and the electromagnetic fields together,
or at least we need to adopt an approximation which allows us to account
for the strong gravity near the black hole. \rev{We closely follow the
results of King \etal\ (1975) and Bi\v{c}\'ak \& Jani\v{s} (1980) who
explored the structure of electromagnetic fields around rotating black
holes and calculated the magnetic flux across their horizon.}

The similarity between the problem of a rotating magnetized body treated
in the framework of classical electrodynamics and the corresponding
black-hole electrodynamics has been widely discussed in the literature
(Thorne \etal\ 1986, chapt.~4). On the one hand, the black-hole problem
is more complex because we have to consider the effects of general
relativity which cannot be ignored when gravity enters into the game,
but, on the other hand, the adopted spacetime represents a vacuum
solution (\rev{more precisely, an electro-vacuum solution}) and it is
thus idealized in this respect. \rev{Otherwise}, rather intricate
relations would be needed in order to determine material properties
(resistivity, polarization, etc.) of an ordinary medium, e.g.\ a plasma,
but all of this is simply and uniquely given in the case of black holes,
when no other matter is present in terms of fluids and solid bodies.
Under astrophysically realistic conditions, the evolution of non-vacuum
(coupled to plasma) electromagnetic fields can be treated by numerical
techniques only. These represent a very broad subject of current
research (for an overview of recent progresses, see Miyama \etal\ 1999,
chapt.\ 4--5). We can thus concentrate ourselves on the effects of
gravity acting on the electromagnetic field.

\section{Motivations}
\subsection{Magnetic fields around astronomical bodies}
Electromagnetic fields exist everywhere in the cosmos and different
approaches have been adopted to determine their intensities (Asseo and
Sol 1987; Kronberg 1994). These fields are usually very weak but they
get amplified inside gravitating bodies such as stars and nuclei of
galaxies during their formation and subsequent evolution. Matter is
attracted and compressed by gravity, and, since it consists partly of
electrically highly conducting plasmas (ionized gases or
electron-positron pairs), the magnetic fields are dragged along and
strengthened. Processes involving complicated and often turbulent plasma
motions are however too complicated to be described in full generality.
Astronomers grasp only some parts of the story called astrophysical
fluid dynamics.

In certain astronomical objects, the magnetic fields are very strong
indeed. It has been deduced from pulsars spin-down rates and also from
direct measurements that the magnetic field of neutron stars reaches
$10^{8}$ tesla (Murakami 1988; Mihara 1990). Recent works about
magnetars (hypothetic neutron stars with super-strong magnetic fields)
suggest even higher values, of the order of $10^{11}$ tesla (Kouveliotou
1998). The existence of magnetars has not been firmly proven yet, and it
is the subject of much debate (Duncan \& Thompson 1992; Zhang \& Harding
2000), however, no basic principle prevents the existence of so
enormously strong magnetic fields in nature. Magnetic fields are limited
in strength only by quantum theory effects: For example, it has been
noticed in extremely magnetized rotators that the energy of the magnetic
field can be converted into gamma rays, but such mechanisms require more
than $10^{12}$ tesla; we will not discuss the regime of such
super-strong magnetic fields here.

The generation of magnetic fields goes hand in hand with creation of
corresponding electric fields which must arise in moving media and, for
that matter, when rotating bodies are involved. For example, electric
fields around magnetized neutron stars ($\approx10^{13}$\,V/m) can
easily accelerate charged particles to speeds comparable with the speed
of light. This effect provides the basis for the pulsar mechanism as it
was proposed already in the early discovery days (Goldreich and Julian
1969). Even before the discovery of pulsars (Hewish \etal\ 1968) it was
proposed by Franco Pacini (1967) that the electric fields due to fast
rotation of the magnetized neutron star may provide the source of energy
in supernova remnants. (For a recent review of these topics, see Michel
\& Li 1999).

\subsection{The interaction of electromagnetic and gravitational fields}
Maxwell's equations are in no way coupled with Newton's gravitational
law, therefore, one could conclude that there is no direct relation
between the electromagnetic field of the body and its gravity. Not so in
general relativity. Energy density of the electromagnetic field, as of
any other field, stands explicitly in Einstein's equations for the
spacetime structure, contributing there as a source of the gravitational
field. Einstein's and Maxwell's equations must be thus considered
simultaneously, and this makes the whole system of equations enormously
difficult to solve in most situations.
                                                        
Fortunately enough, the energy density contained in realistic
electromagnetic fields turns out to be far too low to influence the
spacetime noticeably. Test-field solutions are adequate \rev{for
describing weak electromagnetic fields} under such circumstances, while
the corresponding exact solutions of the coupled Einstein-Maxwell
equations are mainly of academic interest. Nonetheless, this topic is
\rev{suggestive and non-trivial because the relation is often very
subtle between the former case, describing {\em{}linearized
test-field solutions}, and the latter one with strongly
{\em{}nonlinear exact solutions\/} of the coupled equations.}
Notice that, in absolute value, the energy density of a $10^{8}$ tesla
magnetic field sounds quite impressive to those of us who are accustomed
to the usual conditions which can be met on the Earth. On neutron stars,
the energy density of their magnetic fields corresponds to the matter
density of the order of 1~kg/cm$^3$.

By a test-field solution one means the solution of Maxwell's equations in
a fixed spacetime whose structure has been pre-determined by solving
Einstein's equations with no electromagnetic field being present.
Maxwell's equations are solved afterwards and, by assumption, they have
no impact on the gravitational field. This is also the approximation which
we adopt in our present discussion. Namely, we assume the spacetime of a
rotating black hole (the Kerr spacetime; Misner \etal\ 1973,
\rev{chapt.~33}). A reader who is unfamiliar with the language of general
relativity can still grasp the essence of our text if he recognizes that
Maxwell's equations retain their form as if no gravity were present but
an accelerated system were considered instead. Transformation to such a
system with curvilinear coordinates also introduces components of the
metric tensor in Maxwell's equations. Here, the explicit form of the
metric is the Kerr solution (\rev{without charge}; see the Appendix for
mathematical details).

\begin{figure}[t]
\epsfxsize=0.6\hsize
\hspace*{0.16\hsize}
\epsfbox{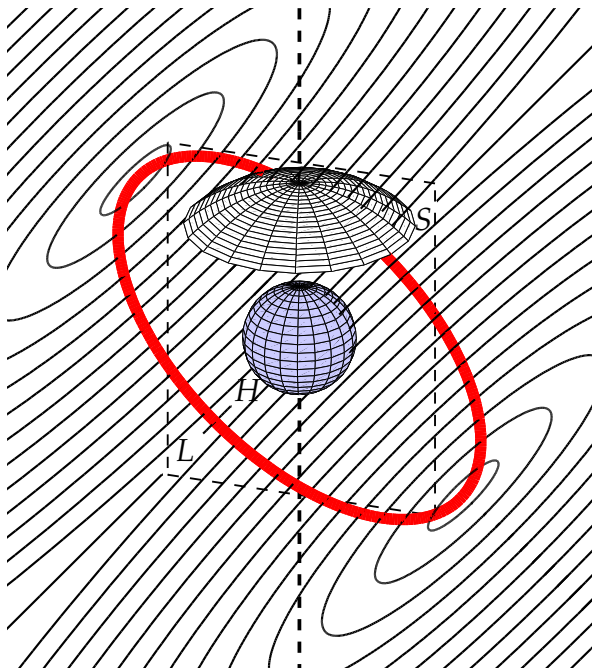}
 \caption{A sketch of the geometrical situation which is discussed in
 this paper: the black-hole horizon is represented by the sphere $H$ around
 which the magnetic field is generated by remote sources, e.g.\ the
 current loop $L$. It resembles an asymptotically uniform field within
 some restricted volume (indicated by the rectangle) around the black
 hole. In general, the magnetic field can be inclined with respect to
 the rotation axis of the hole (dashed line). The shape of the field
 lines and the corresponding magnetic flux across the surface $S$ (or
 any other space-like surface) can be easily determined if
 the direction of the field coincides with the rotation axis of the hole
 (an aligned field). But with non-aligned fields and very near the
 horizon the situation becomes much more complicated.
 \label{f0}}
\end{figure}

We consider two different situations hereafter: first, the fields
parallel to the rotation axis of the black hole (aligned fields), and
then those which are oblique. While the aligned fields are not wound up
around the hole (there is no plasma which could induce a toroidal
component of the field), the oblique fields are indeed dragged due to
the hole's gravomagnetic influence. The two cases are thus qualitatively
different. The main simplification of the former one stems from the fact
that the aligned electromagnetic field obeys identical symmetries as the
gravitational field.

\section{Field lines and fluxes}
The general definition of the magnetic and the electric fluxes should
reflect the intuitive idea that the flux amounts to the number of field
lines crossing the unit surface area at perpendicular orientation,
taking into account the observer dependence of the field lines. Since
the flux is a scalar quantity, it enables us to define surfaces of
constant flux in appropriate way with the corresponding field lines
lieing within these surfaces.
                   
Let us consider an arbitrary two-dimensional \rev{spacelike} surface
$\calS$ around the black hole (Figure~\ref{f0}). Such a surface
can represent, e.g., a part of a constant radius sphere in spheroidal
coordinates $r$, $\theta$, $\phi$. The magnetic and the electric fluxes
across the surface, $\Phim$ and $\Phie$ respectively, are defined by
\beq
\Phim=\int_{\calS}\bm{F$\wedge$d}{\calS},\quad
\Phie=\int_{\calS}\bm{$^*\!$F$\wedge$d}{\calS},
\label{flux}
\eeq
where $\bm{F}$ denotes the electromagnetic-field 2-form, $\bm{$^*\!$F}$
is its \rev{Hodge-dual with respect to the spacetime metric, and
$\bm{d}\calS$ refers to the element of surface embedded in the
four-dimensional spacetime. The symbol $\wedge$ denotes the wedge
product of the two forms; see \S2 of Bi\v{c}\'ak \& Jani\v{s} (1980)
for further details.} In particular, if $\calS$ is a part of the
$r={\rm{const}}$ sphere, then
$\Phim=\int{}F_{\theta\phi}\,\rd\theta\,\rd\phi$, which is a natural
formula for the magnetic flux where $F_{\theta\phi}$ has the meaning of
the radial component of the magnetic field. In the case of a closed
surface $\calS$, by applying the Gauss theorem, the integral $\Phim$
must vanish (no magnetic charges) while the integral $\Phie$ is then
proportional to the total electric charge contained in the volume
enclosed by $\calS$.

Maxwell's equations in vacuum can be written in compact form
\beq
\bm{dF}=0, \quad \bm{d$^*\!$F}=0,
\label{mx}
\eeq
where the operator $\bm{d}$ denotes exterior differentiation in 
curved spacetime. The general form of the solution of eqs.\ (\ref{mx})
can be expressed only in terms of infinite expansions (sum of the
multipoles), subject to boundary conditions, but explicit expressions
have been found for special cases, namely, the asymptotically uniform
magnetic field (King \etal 1975). Even such a highly simplified solution
becomes rather complicated when the field is not aligned exactly with
the rotation axis of the hole, as we illustrate below.

Now we employ the above-given definitions and introduce the concept of
surfaces of constant flux. For simplicity, let us start with the fields
which are axially symmetric. In this case the black hole behaves like an
aligned rotator: once it is placed in the uniform magnetic field, the
hole distorts the original structure of the magnetic field and it
induces electric fields.

\subsection{Axisymmetric fields}

\begin{figure*}[t]
\hspace*{0.16\hsize}
\epsfxsize=0.39\hsize
\makebox[0.4\hsize]{\hspace*{-3ex}\epsfbox{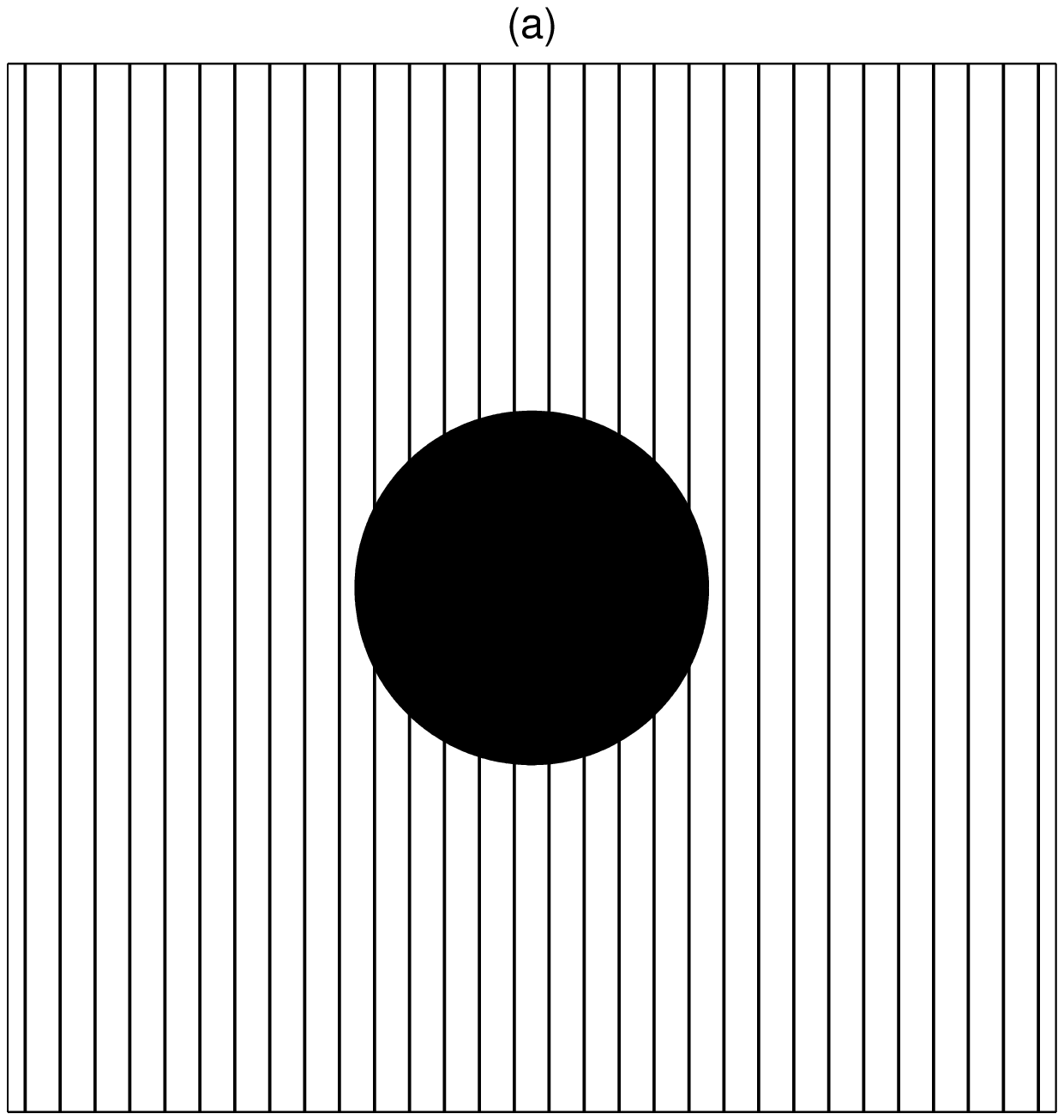}}
\epsfxsize=0.39\hsize
\makebox[0.4\hsize]{\hspace*{3ex}\epsfbox{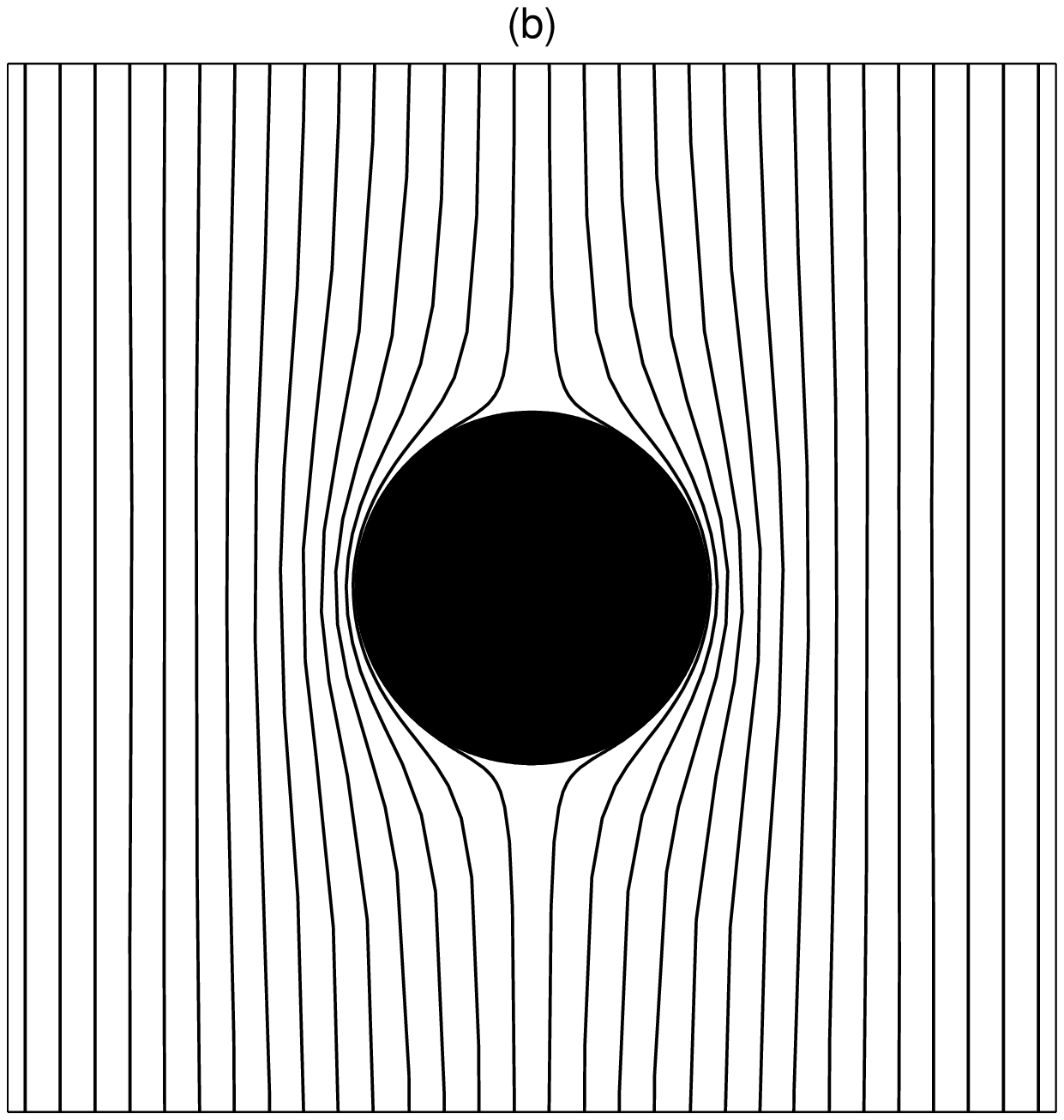}}
 \caption{Surfaces of constant flux $\Phim$ for the axisymmetric
 magnetic field. Here we plot the intersections of these cylindrical
 surfaces with the azimuthal plane $\phi={\rm{const}}$. The black-hole
 horizon is denoted by a circle, the rotation axis is vertical. Two
 cases are shown: (a)~$a=0$ (a static black hole), and (b)~$a=M$ (a
 maximally rotating black hole). Notice that rapidly rotating holes
 exhibit the Meissner-type effect (expulsion of the magnetic field out
 of super-conducting bodies). Only a small region is shown near the
 horizon, where the effect of the black hole is apparent.
 \label{f1}}
\end{figure*}

The gravitational field of a black hole in steady rotation possesses
two evident symmetries which are crucial in our discussion. These are
stationarity and axial symmetry. As a consequence of the symmetries one
can employ spheroidal coordinates in which the horizon of the hole is
located at constant radius, $r=r_+$, and all partial derivatives with
respect to azimuth $\phi$ and time $t$ vanish. Due to rotation, however,
the gravitational field is not spherical and it depends on the latitudinal
coordinate $\theta$ (see the Appendix). The radius of the horizon is given
in terms of the black hole's mass $M$ and its specific angular momentum
$a$: $r_+=M+\surd(M^2-a^2)$. Rotation of the hole cannot be arbitrarily
large and it is limited by the maximum value of the angular momentum,
$a=M$ in geometrized units.\footnote{Here we use geometrized units ($c=G=1$) 
which are often employed in relativity. In physical units (denoted by a
tilde), $r_+=\tilde{r}_+=\tilde{G}\tilde{c}^{-2}[\tilde{M}+
\surd(\tilde{M}^2-\tilde{c}^2\tilde{a}^2/\tilde{G}^2)]$, where
$\tilde{c}$ is the speed of light and $\tilde{G}$ is the gravitational
constant.} 

The electromagnetic field may or may not possess the same symmetries
as the gravitational field. Naturally, the problem is greatly simplified
by assuming axial symmetry and stationarity for both fields. We start by
examining the asymptotically (i.e.\ far from the hole) uniform magnetic field
which is aligned with the rotation axis and satisfies the required symmetries
automatically (Wald 1974). Let us specify the surface $\calS$ as part of
the sphere $r={\rm{}const}\geq{}r_+$ with the circular boundary given by
$\theta={\rm{}const}$ (Fig.~\ref{f0}). Using eq.\ (\ref{flux}) and the
field components from the Appendix one can calculate the fluxes.
Denoting the asymptotical strength of the magnetic field by
$B_{\parallel}$ we find for the magnetic flux
$\Phim(r,\theta;a,M,B_{\parallel})$ across the cap $\calS$:
\beq
\Phim=\pi{B_{\parallel}}\left[\Delta+
 \frac{2Mr}{\Sigma}\left(r^2-a^2\right)\right]\sin^2\theta,
 \label{phim}
\eeq
where $\Delta=r^{2}-2Mr+a^{2}$, $\Sigma=r^{2}+a^{2}\cos^{2}\theta$.
Furthermore, setting $\Phim=C={\rm{}const}$, and solving for
$r{\equiv}r_{_C}(\theta)$ one finds a surface of constant
flux. \rev{Eq.~(\protect\ref{phim}) with $r=r_+$,
$\theta=\pi/2$ can be reduced to eq.~(27) of Bi\v{c}\'ak \& Jani\v{s}
(1980) for the flux of an aligned magnetic field across the hemisphere.}

A set of constant-flux surfaces can be parametrized by the values of
$C$, where each of the surfaces forms a flux tube with topology of a
cylinder along the rotation axis. In other words, \rev{$r_{_C}(\theta)$
determines the poloidal shape of the flux-tube boundary (i.e.\ its
intersection with a $\phi={\rm{const}}$ plane).} In particular, one can
start by calculating the flux across the whole hemisphere (i.e.\ one
half of the surface of the horizon, $r=r_+$ and $\theta\leq\pi/2$), and
then, using the same value of $\Phim(r_+,\pi/2)\equiv\Phi_{\rm{}max}$
one proceeds to higher $r$'s up to spatial infinity,
$r\rightarrow\infty$. In this way one finds the effective
cross-sectional area for the capture of the magnetic flux which is a
circle of radius $2M\surd[1-a^2/(Mr_+)]$. This circle is a section of
the magnetic flux tube which comes from spatial infinity
($r\rightarrow\infty$, $\sin\theta\rightarrow0$, where the field is
strictly uniform) and eventually crosses the black-hole horizon.

In the non-rotating case, the cross-section is just a circle of radius
$r_+(a=0)=2M$ while the maximum flux is
$\Phi_{\rm{}max}={\pi}B_{\parallel}{r_+^2}$. But the flux across the
hole decreases with $a$ increasing, and in the maximally rotating case
($a=M$) the field becomes expelled out of the hole completely
(Bi\v{c}\'ak \& Jani\v{s} 1980). In other words, the cross-section
shrinks to zero. This conclusion corresponds to $\Phim(r=r_+,a=M)=0$
[cf.\ eq.~(\ref{phim})], and in this sense the horizon of the maximally
rotating black hole bears properties of superconducting bodies, which do
not permit external magnetic fields to thread their surfaces.

Two special cases \rev{are shown in} Figure~\ref{f1}. First, for a
non-rotating black hole ($a=0$), the surfaces of constant flux appear
indeed as cylinders, $r\sin\theta={\rm{const}}$. The exact shape of
cylindrical surfaces of course depends on the choice of the radial
coordinate, which is not unique. Unlike angular coordinates which are
defined straightforwardly, one could \rev{make a transformation}
$R\equiv{R(r)}$. Such a transformation would introduce a distortion to
the lines which came out straight in Fig.~\protect\ref{f1}a. Radius $r$
has a clear meaning far from the black hole where the spacetime becomes
flat and $r$ is simply the spherical radial coordinate.

Now we consider the magnetic field lines which, by definition, lie in
the surfaces of constant flux and determine the direction of Lorentz
force acting on charged particles. Such lines of force are observer
dependent (the Lorentz force depends on the particle's velocity). One
thus has to be careful in the choice of preferred observers whose motion
defines the lines that are plotted in graphs. In practice one often
chooses a frame connected to observers matching in some natural manner
the spacetime properties. For example, one can build the frame by
requiring that it corotates with the black hole around its axis at
such a velocity that the azimuthal component $p_\phi$ of energy-momentum
four-vector is zero ($\ell=p_\phi$ \rev{is constant of motion, hence the
name for observers} attached to this frame: zero angular momentum
observers). Another possibility would be to connect the frame with
photons ingoing into the hole. We will explore both options hereafter.

\begin{figure*}[t]
\hspace*{0.16\hsize}
\epsfxsize=0.39\hsize
\makebox[0.4\hsize]{\hspace*{-3ex}\epsfbox{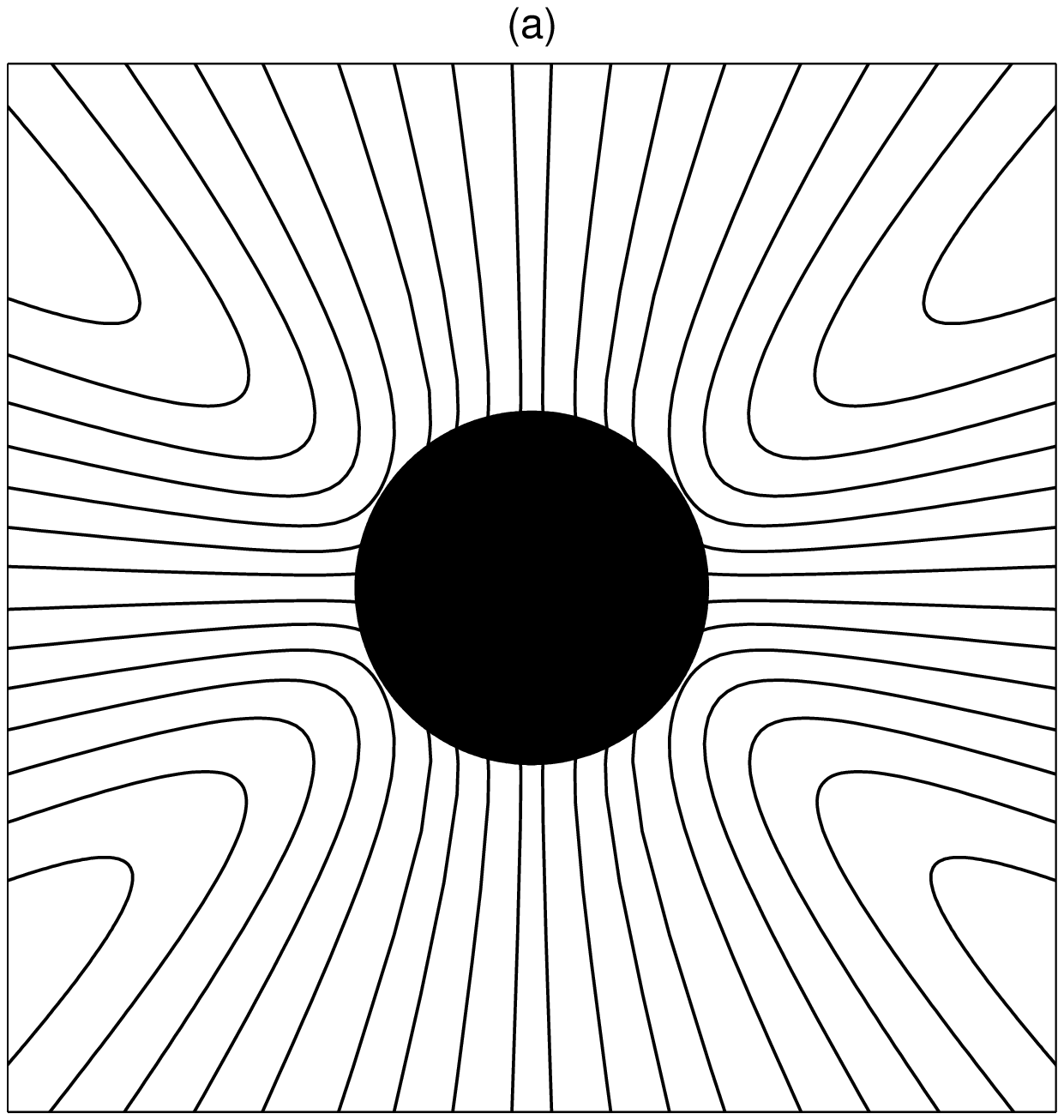}}
\epsfxsize=0.39\hsize
\makebox[0.4\hsize]{\hspace*{3ex}\epsfbox{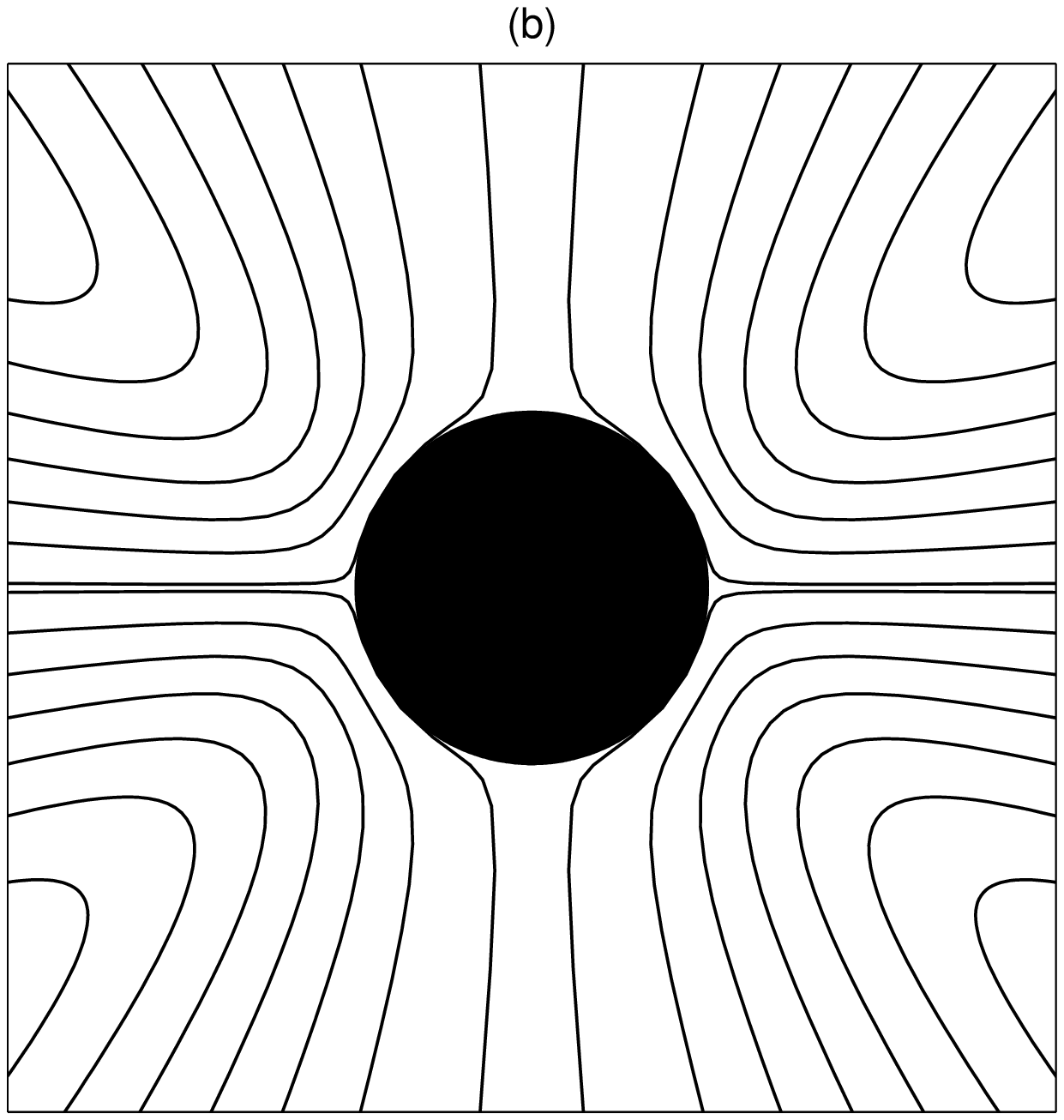}}
 \caption{The electric field can be induced by the rotating black hole.
 In this figure, electric lines of force are plotted corresponding
 (a)~to the uniform aligned magnetic field near the fast rotating black
 hole ($a=0.95M$); (b)~to a similar situation near the maximally
 rotating hole ($a=M$, as in Fig.~\protect\ref{f1}b).
 \label{f2}}
\end{figure*}

With aligned fields our task is quite simple (Christodoulou
\& Ruffini 1973). The lines indicating the shape of the constant flux
surfaces in Fig.~\ref{f1} are parallel to the magnetic force acting on
(hypothetical) magnetic monopoles connected with zero angular momentum
observers. This can be verified directly from the expression for the
acceleration of a particle due to the Lorentz force,
\beq
m\dot{\bm{u}}=q_{\rm{m}}\bm{$^*\!$F.u},
\label{lorentz}
\eeq
where on the left-hand side the dot denotes total derivative with
respect to particle's proper time, $m$ and $q_{\rm{m}}$ are its mass
and magnetic charge respectively; $\bm{u}$ is the four-velocity of the
inertial reference frame coinciding at each point with the frame of
zero angular momentum observers (the first choice of preferred
observers mentioned above). The field lines are given by the ordinary
differential equation
\beq
\frac{{\rd{r}}}{{\rd\theta}}=\frac{B_r}{B_\theta},
\label{l1}
\eeq
where $B_r={^*\!F}_{r\nu}\,u^{\nu}$, $B_\theta={^*\!F}_{\theta\nu}\,u^{\nu}$
are the magnetic field intensity \rev{components}
calculated by projecting the tensor
${^*\!F}_{\mu\nu}$ onto \rev{the observer's} four-velocity.

The electric fluxes and field lines can be introduced in a similar
manner; one only needs to interchange the electromagnetic field tensor
by its dual, the magnetic charge by the electric charge, and vice versa
wherever they appear in the above-given formulae. The electric field
induced by rotation of the hole is shown in Figure~\ref{f2}. (The black
hole could have also its own electric charge, giving rise to the Coulomb
field which would then have to be superposed to the induced
field; but this is just a minor complication.) It is evident that the
induced electric field vanishes in the non-rotating case, but quite
unapparent is the form of the induced field in Fig.~\ref{f2},
almost radial far from the hole. Based on the classical analogy with a
rotating sphere, one would perhaps expect a quadrupole-type component,
but here the leading term of the electric field arises due to
gravomagnetic interaction which is a purely general-relativistic effect
(Ciufolini \& Wheeler 1995). This electric field falls off radially as
$r^{-2}$.

\subsection{Oblique fields}

Now we explore the magnetic field which is \rev{not aligned} with the
rotation axis, though it still remains asymptotically uniform
(Bi\v{c}\'ak \& Jani\v{s} 1980). Let us recall that we are dealing with
the test-field solution (linearized in the field intensity) which can be
written as a sum of two parts: the first one conveys the contribution of
the aligned component as introduced in the previous section
(proportional to $B_{\parallel}$), and the second one corresponds to
more complicated asymptotically perpendicular field
(${\propto}B_\perp$). We turn our attention to this second part.

\begin{figure*}[t]
\hspace*{0.16\hsize}
\epsfxsize=0.39\hsize
\makebox[0.4\hsize]{\hspace*{-3ex}\epsfbox{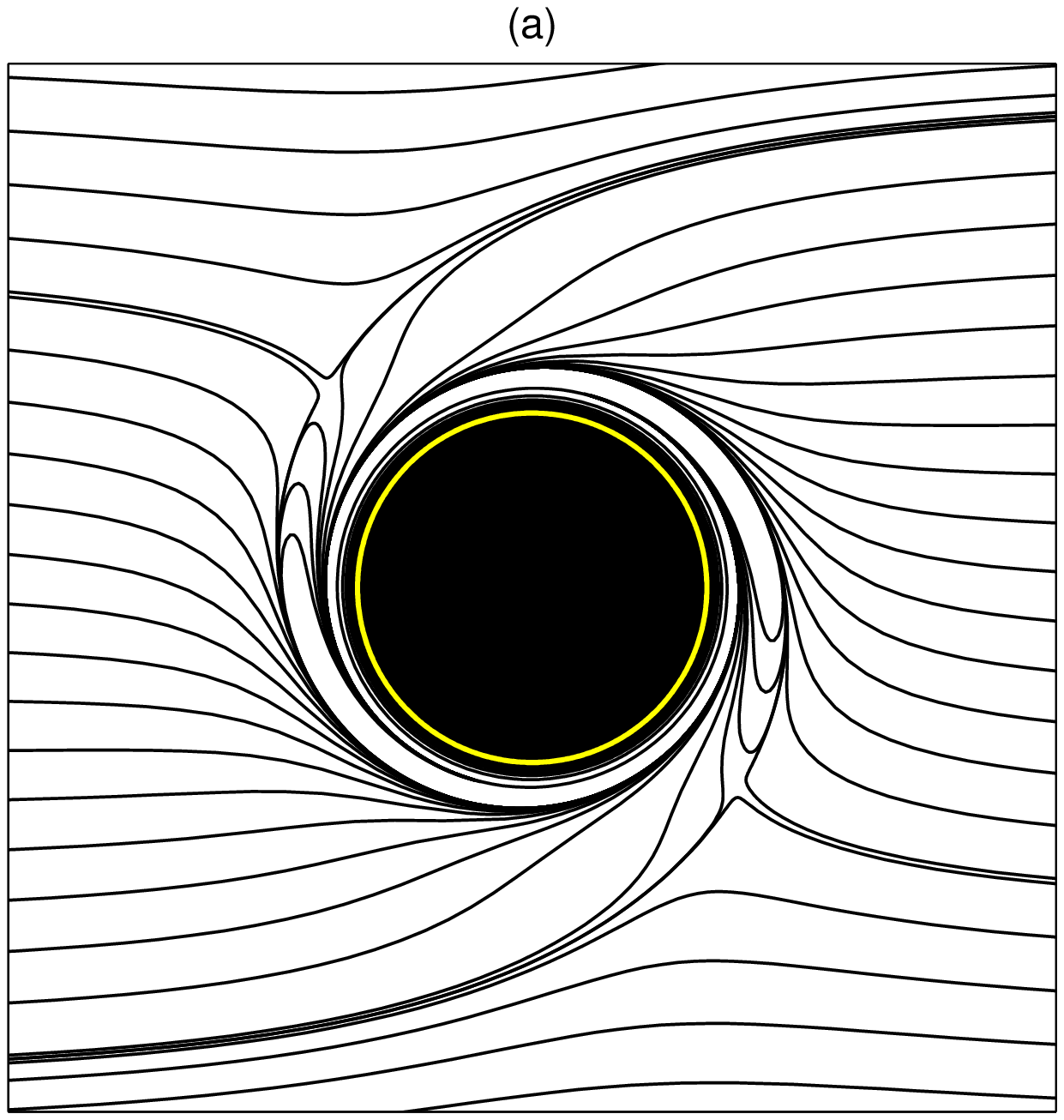}}
\epsfxsize=0.39\hsize
\makebox[0.4\hsize]{\hspace*{3ex}\epsfbox{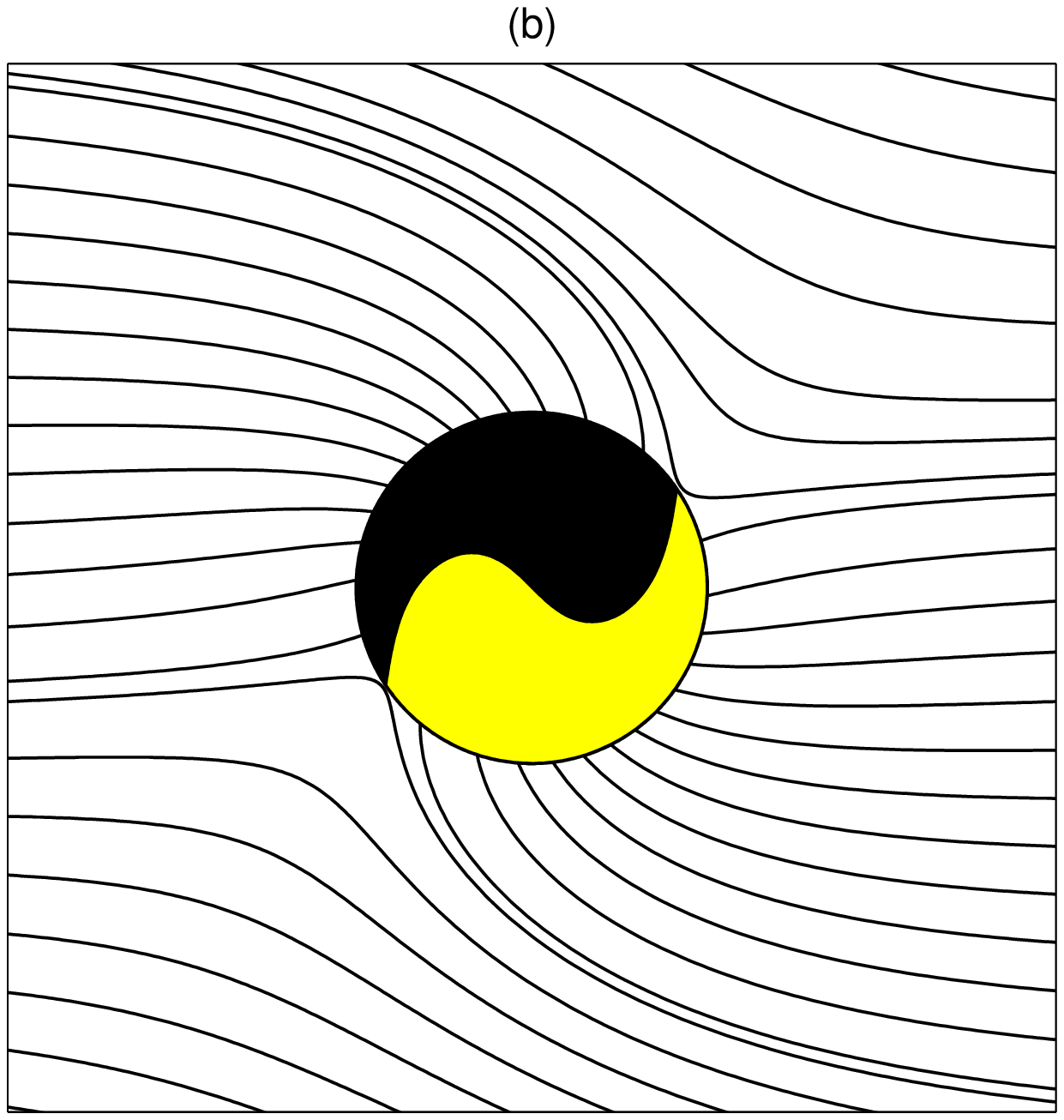}}
 \caption{Lines of the magnetic field which is asymptotically uniform
 and perpendicular to the rotation axis. The 
 equatorial plane is shown as viewed
 from top, i.e.\ along the rotation axis, (a)~in the frame of zero angular
 momentum observers orbiting at constant radius; (b)~in the frame of
 freely falling observers. In the panel (b), two regions of
 ingoing/outgoing lines are distinguished by different levels of shading
 of the horizon. The hole rotates counter-clockwise ($a=M$).
\label{f3}}
\end{figure*}

The lines of asymptotically perpendicular field do not reside in one
plane and it is not easy to visualize them. The equatorial plane
$\theta=\pi/2$ is an exception (for obvious reasons of symmetry) and we
thus start here by plotting the lines which are defined in a manner
analogous to eq.~(\ref{l1}):
\beq
\frac{\rd{r}}{\rd\phi}=\frac{B_r}{B_\phi}.
\label{l2}
\eeq
The lines remain in the equatorial plane ($B_\theta=0$ there) but they
do not look straight at all. Instead, the lines are wound up by rotation
of the hole whenever $a\neq0$ (Figure~\ref{f3}a), and these dragging
effects are progressively enhanced near horizon. The reason of this
behaviour is easy to understand, and it is related to the choice of the
reference frame. Zero angular momentum observers, to which we attach
the frame, circulate around the hole at constant radius; therefore, they
experience a rapidly increasing gravitational pull when
$r{\rightarrow}r_+$ and must be supported by force which eventually
diverges at $r=r_+$. There the frame becomes nonphysical in the sense
that it cannot be realized by any kind of physical particles. A question
arises naturally: Can we go over to another frame which does not suffer
from such a drawback? This new frame should remain physical everywhere
outside the black hole, including the horizon itself.

Let us now transform the field to the frame which is connected to
photons falling into the hole straight along latitudinal rays $\theta$
constant. In other words, the infall is simply radial at large distance.
Such a frame must remain physical also on the horizon by its very
definition: it is realized with the help of particles, motion of which
is governed by the hole's gravity (we could use with similar results
also particles of nonzero rest mass instead of photons).
Figure~\ref{f3}b shows the magnetic field lines constructed in this
frame, sometimes called the Kerr ingoing frame (cf.\ Appendix). It is
evident from the figure that the observers do not experience any
particular effects related to the magnetic field structure in the
infalling frame near the horizon.

Besides the field lines, we have indicated (by
light shading) that part of the horizon where the field enters the hole
(the radial component of the local magnetic field is negative there).
Integrating over this area one obtains the total ingoing magnetic flux,
$\Phi_{\rm{}max}$, which again depends on the rotation parameter $a$,
but now in a way substantially different than it was for the aligned
fields. We could not derive an analytical formula for the non-aligned
flux but we calculated numerically that, e.g., for $a=0.95\,M$
the maximum flux reaches $\Phi_{\rm{}max}\dot{=}5.1B_{\perp}r_+^2$. This
value is {\em{}larger\/} than the corresponding
$\Phi_{\rm{}max}={\pi}B_{\parallel}r_+^2$ in the aligned case. Moreover,
the ingoing flux does not vanish even when $a=M$.

\rev{We remark that our total ingoing flux $\Phi_{\rm{}max}$ exceeds
the maximum flux across a hemisphere,
$\tilde{\Phi}_{\rm{}max}=B_\perp{\pi}r_+^2\surd[1+a^2(1+r_+)^2/r_+^4]$.
(The latter formula for $\tilde{\Phi}_{\rm{}max}$ corresponds to
eq.~(33) of Bi\v{c}\'ak \& Jani\v{s} (1980) after correcting their term
under the square root.) The reason of the difference between
$\Phi_{\rm{}max}$ and $\tilde{\Phi}_{\rm{}max}$ is evident from
Fig.~\ref{f3}b: the shaded surface where the flux enters into the
horizon does not coincide in shape with a hemisphere. It is
$\Phi_{\rm{}max}$ which determines the flux connecting the horizon to
distant regions outside the hole, while $\tilde{\Phi}_{\rm{}max}$ is not
suitable for this purpose because an inclined hemisphere on the rotating
black-hole horizon always contains the regions of both ingoing and
outgoing field lines. However, the difference is very small between the
two quantities.}

\rev{By Maxwell equations} the ingoing flux is balanced by the outgoing one,
so that the total magnetic charge of the hole vanishes. What appears
less obvious is the fact that, for some observers, there exist closed
loops of magnetic lines which start from the horizon, make a turn, and
plunge back into the horizon. This is not the case of the infalling
frame, but it can be observed by zero angular momentum observers near the
horizon (see Fig.~\ref{f3}a). For them the magnetic flux connecting the
hole with distant spatial regions is somewhat less than the maximum
ingoing flux on the horizon. Let us note that the field lines connecting
the hole with distant regions are relevant for astrophysical problems,
since they can establish a link with surrounding plasma and transfer
torques between the hole and the plasma.

\begin{figure}[t]
\hspace*{0.16\hsize}
 \epsfxsize=0.84\hsize\epsfbox{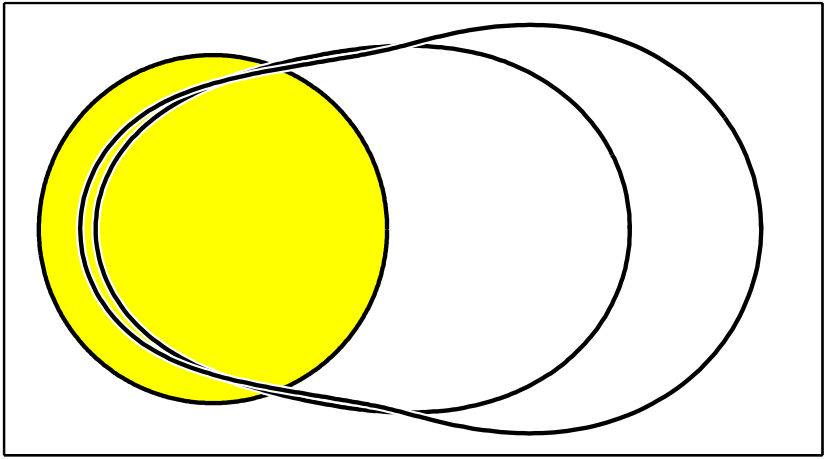}
 \caption{Cross-sectional area for the capture of magnetic field lines
 (asymptotically uniform magnetic field perpendicular to the rotation axis).
 The three curves correspond to different values of the black-hole angular
 momentum: $a=0$ (cross-section is the perfect circle; its projection
 coincides with the black-hole horizon, indicated here by shading),
 $a=0.95\,M$, and $a=M$ (the most deformed shape refers to the
 maximally rotating case).
 \label{f4}}
\end{figure}

At this point a comment is worth regarding the choice of the two frames
which we have employed. Although there is an infinite number of possible
choices, depending on the motion of preferred observers to which the
frame is connected, the main difference stems from the fact that
$r={\rm{const}}$ observers are kept above the black hole with the help
of some supporting force, e.g.\ by using a thruster rocket (even very
close to horizon where the required support increases tremendously)
while freely moving observers are allowed to fall in.

\rev{The capture of the magnetic flux by the black hole can be
characterized by the cross-sectional area also in the case of oblique
magnetic fields. The cross-section is defined in much the same way as
for the aligned field (see previous section). However, when deriving
eq.~(\ref{phim}) we took the advantage of axial symmetry: each flux tube
of the aligned field had a circular cross-section, while for non-aligned
fields the sections are deformed, and their shape remains unknown until
the field lines are found.}

Hence, for the asymptotically perpendicular field we could not find the
shape of flux tubes analytically, although the field lines can still be
traced numerically even outside the equatorial plane. \rev{One finds the
flux tube defining the cross section by shooting from spatial infinity
the field lines with a given inclination to the hole's axis of rotation.
Numerically constructed lines were used to determine the cross-sectional
area of the hole for the capture of the non-aligned magnetic field.} In
this way the effective cross-section is obtained in Figure~\ref{f4} by
tracing in a fine grid those field lines that are just on the verge
between hitting the hole and missing it. We used the magnetic lines in
the ingoing frame to construct this plot. Notice how the cross-sectional
area is deformed and enlarged by increasing the angular momentum of the
hole.

\section{Conclusions}
We explored the surfaces of constant magnetic flux and the field lines
of the uniform magnetic field near rotating black holes. The test-field
solution which we used is simple enough to exhibit the frame-dragging
effects in a crystal-clear form. In addition to that, one can show, by
expanding the field around the hole into multipoles, that the leading
component (the one which dominates near the horizon) corresponds just to
this asymptotically uniform field. We observed how the magnetic field in
vacuum is dragged and twisted by pure gravitational effects of the hole.
When plasma is also present, this effect can lead to rapid reconnection
of the field lines, resulting thus in flaring and explosive releases of
energy (Mestel 1999). But even the vacuum fields are relevant in
astrophysics. It has been suggested (Tomimatsu 2000) that they can act
as seeds for the dynamo process, which subsequently amplifies the
magnetic fields in rotating magnetospheres of compact objects.

\rev{The possibility of extracting the rotational energy via the magnetic
field requires that the field threads the horizon. Bi\v{c}\'ak \& Jani\v{s} 
(1980) remarked that the efficiency of this process may be seriously
limited by the magnetic-field expulsion if the hole rotates very rapidly, 
but this constraint applies only to the field which is uniform and aligned.}
As we illustrated in previous section, non-aligned fields thread the horizon 
even in the case of extreme rotation, and the complicated structure of the
field suggests that significant acceleration of particles takes place
due to reconnection of the field lines near the horizon.

The problem of magnetic fluxes threading black holes has also other
astrophysically relevant implications. The fields around rotating holes
are of particular interest because the newly born holes probably gain a
lot of angular momentum at the moment of creation by violent contraction.
Magnetic fields could extract rotational energy from the black hole back
by converting it to the outflowing Poynting flux and to kinetic energy
of accelerated plasma flows (Blandford \& Znajek 1977; for recent
accounts of the problem, see Punsly 1996; Ghosh \& Abramowicz 1997).
Indeed, such outflows are observed, but the regions where they originate
remain below resolution capabilities of present-day techniques. 

As still another exciting application, it has been proposed that the
energy of the electromagnetic field is violently released in some transient
types of astronomical objects. The process could start
if a massive torus is formed as a relic of a close encounter between two
neutron stars, one of them being disrupted by tidal forces exerted by
its companion. The orbiting matter then undergoes the gravitational
collapse into the black hole due to a certain kind of rapid instability
(Daigne \& Mochkovitch 1997; Abramowicz \etal\ 1998). If the magnetic
field were initially held by the torus (as we sketched in Fig.~\ref{f0})
then its collapse would increase the magnetic intensity tremendously (due to
the flux conservation) while subsequent disappearance of the torus leads
to the lack of this support and to sudden expansion of the field lines.
Rapid acceleration of the surrounding plasma results thereof, and
eventually a burst of radiation is produced. Similar processes have been
proposed as a possible origin of gamma-ray bursts (Hanami 1997; Daigne
\& Mochkovitch 2000; Lee \etal\ 2000), powerful and as yet unexplained
events which are routinely detected by specialized satellites.

{\protect\small
\ack
The authors thank the referee for helpful comments which helped
us to improve our text. MD and VK acknowledge support from the grants
GA\,CR 205/00/1685 and GA\,UK 63/98 in the Czech Republic.}

\setcounter{equation}{0}
\renewcommand{\theequation}{A.\arabic{equation}}

\section*{Appendix}
\rev{Gravitational field of a rotating black hole is described by Kerr
metric. In this Appendix we summarize the relevant expressions for the
spacetime metric, for the structure of weak electromagnetic fields around
the hole, and for the electric and the magnetic lines.}

\subsection*{The spacetime metric and the electromagnetic field}
Kerr metric has the form (Misner \etal\ 1973, \rev{chapt.~33})
\begin{equation}
\fl {\rm d}s^{2} =
  -\Delta\Sigma{{\cal A}^{-1}}\,{\rd}t^{2}
   +\Sigma\Delta^{-1}\,{\rd}r^{2}
   +\Sigma\,{\rd}\theta^{2}
   +{\cal A}\Sigma^{-1}\sin^{2}\theta\;
   \left({\rm{d}}\phi-\omega\,{\rd}t\right)^{2};
\end{equation}
here, spheroidal (Boyer-Lindquist) coordinates,
$x^\mu\equiv\{t,r,\theta,\phi\}$, and geometrized units ($c=G=1$) have
been used, $M$ and $a$ denote the mass and the specific angular momentum
of the body, $\Delta(r)=r^{2}-2Mr+a^{2}$,
$\Sigma(r,\theta)=r^{2}+a^{2}\cos^{2}\theta$,
${\cal A}(r,\theta)=(r^{2}+a^{2})^{2}-{\Delta}a^{2}\sin^{2}\theta$, and 
$\omega(r,\theta)=2Mar/\cal{ A}$.
Notice that on the horizon $\Delta(r_+)=0$.

The electromagnetic field tensor $\bm{F}$ can be expressed in terms
of the four-potential, $\bm{F}=\bm{dA}$. Spelling out the components
explicitly, one obtains
\beq
F_{\mu\nu}=\frac{\partial{A_\nu}}{\partial{x^\mu}}-
\frac{\partial{A_\mu}}{\partial{x^\nu}},
\eeq
where (\rev{eq.~(A4) of} Bi\v{c}\'ak \& Jani\v{s} 1980)
\begin{eqnarray*}
\fl
A_t =
 B_{\parallel}a\left[Mr\Sigma^{-1}\left(1+\cos^2\theta\right)
 -1\right] + B_{\perp}aM\Sigma^{-1}\Psi\sin\theta\cos\theta, \\
\fl
A_r = -B_{\perp}(r-M)\sin\theta\cos\theta\sin\psi, \\
\fl
A_{\theta}=
 -B_{\perp}\Big[\left(r\sin^2\theta+M\cos^2\theta\right) a\cos\psi
 + \Big(r^2\cos^2\theta+\left(a^2-Mr\right)\cos2\theta\Big)
 \sin\psi\Big], \\
\fl
A_{\phi} =
B_{\parallel}\left[{\textstyle\frac{1}{2}}\left(r^2+a^2\right)
 -a^2Mr\Sigma^{-1}\left(1+\cos^2\theta\right)\right] \sin^2\theta \\
\lo
 - B_{\perp}\left[\Delta\cos\psi+\left(r^2+a^2\right)
 M\Sigma^{-1}\Psi\right] \sin\theta\cos\theta,
\end{eqnarray*}
$\psi=\phi+a\delta^{-1}\ln\left[\left(r-r_+\right)/\left(r-r_-\right)\right]$, 
$\Psi=r\cos\psi-a\sin\psi$, $\delta=r_+-r_-$, $r_{\pm}=M\pm\sqrt{M^2-a^2}$.

\subsection*{The field lines}
\rev{The components of electric and magnetic intensities, $E_{(k)}$
and $B_{(k)}$, measured by a physical observer (i.e., for example,
experienced by a charged particle) can be obtained by projecting
$F_{\mu\nu}$ and its dual onto the local tetrad of basis vectors
$\bm{e}_{(k)}$,}
\begin{eqnarray}
E_{(k)} &=&
 \bm{e}_{(k)}^{\mu}F_{\mu\nu}\,u^{\nu}=F_{(k)(t)},
 \label{intensity1} \\
B_{(k)} &=&
 {\bm{e}_{(k)}^{\mu}}{^*\!F}_{\mu\nu}\,u^{\nu}={^*\!F}_{(k)(t)},
 \label{intensity2}
\end{eqnarray}
where $u^\nu\equiv\bm{e}_{(t)}^{\nu}$ is the observer's four-velocity, 
and the remaining three basis vectors are chosen as space-like, mutually
perpendicular vectors. In particular,
we have used the frame of radially infalling photons. The tetrad
defining this frame is obtained by transforming the time and the
azimuthal angle,
$\{t,\phi\}\rightarrow\{v,\psi\}$, in the following way:
$\bm{d}v = \bm{d}t + \left(r^2+a^2\right)\Delta^{-1}\bm{d}r$, 
$\bm{d}\psi = \bm{d}\phi+ a\Delta^{-1}\bm{d}r$ (Kerr ingoing coordinates).
Here, indices $(k)$ in round parentheses correspond to spatial coordinates
$(r)$, $(\theta)$, $(\phi)$.

Notice that, in an orthonormal tetrad, one can shift spatial indices
of a tensor up and down without any change, while the sign of each 
tensorial quantity must be changed when shifting the time index $(t)$.

Coordinate components of the electric and the magnetic intensities 
(\ref{intensity1})--(\ref{intensity2}) are
\begin{eqnarray}
E^{\mu} & = & {F^\mu}_\nu\,u^\nu  =  {e_{(k)}}^\mu E^{(k)} \,, \\
B^{\mu} & = & {^*\!F^\mu}_\nu\,u^\nu = {e_{(k)}}^\mu B^{(k)} \, .
\end{eqnarray}
Under the Lorentz force, the motion of electrically/magnetically charged
test particles obeys equations
\beq
m\,a^\alpha=q_{\rm{e}}\,{F^\mu}_\nu\,u^\nu,
\quad
m\,a^\alpha=q_{\rm{m}}{^*\!F^\mu}_\nu\,u^\nu,
\eeq
respectively (cp.\ eq.~(\ref{lorentz})). Here, $m$ is the mass of the particle
and $a^\alpha=u^\alpha_{;\beta}\,u^\beta$ is four-acceleration.

The electric and the magnetic lines are determined by the differential
equations
\beq
\frac{{\rm d}x^\mu}{{\rm d}s} = E^\mu,\, \quad
\frac{{\rm d}x^\mu}{{\rm d}s} = B^\mu \, .
\eeq
The shape of the lines depends on the choice of the particular frame and 
coordinates which are used to draw the curves.

\subsection*{Two examples of the tetrads}
\rev{As an example, we can write the explicit form of the tetrads
mentioned in the text (for a useful summary of their properties and for
further details, see Semer\'ak 1993). The tetrad of zero angular
momentum observers is in Boyer-Lindquist coordinates}
\begin{eqnarray}
\bm{e}_{(t)} & = & 
 {\cal A}^{1/2}
 \left(\Delta\Sigma\right)^{-1/2}\,\bm{[}1,0,0,\omega\bm{]}\,,
 \label{zamo1} \\
\bm{e}_{(r)} & = & \bm{[}0,\Delta^{1/2}\Sigma^{-1/2},0,0\bm{]}\,, 
\\
\bm{e}_{(\theta)} & = & \bm{[}0,0,\Sigma^{-1/2},0\bm{]}\,, 
\\
\bm{e}_{(\phi)} & = &
 \bm{[}0,0,0,{\cal A}^{-1/2}\Sigma^{1/2}\sin^{-1}{\!\theta}\bm{]}\,.
 \label{zamo4}
\end{eqnarray}
Observers at rest with respect to the tetrad (\ref{zamo1})--(\ref{zamo4})
follow an orbit with $r$, $\theta$ constant, and with angular momentum
$\ell=0$. Notice that these observers are accelerated rather than
freely falling.

For the tetrad attached to observers at free fall (along
$\theta={\rm{const}}$ from the rest at infinity) one obtains
\begin{eqnarray}
\hat{\bm{e}}_{(t)} & = &
 \Sigma^{-1}\bm{[}1-(r^2+a^2){\cal K},-[2Mr(r^2+a^2)]^{1/2},
 0,-a{\cal K}\bm{]}\,,
\\
\hat{\bm{e}}_{(r)} & = &
 {\cal A}^{1/2}\Sigma^{-1}\,\bm{[}1-{\cal K},1,0,
 a{\cal A}^{-1}\Sigma+\omega(1-{\cal K})\bm{]}\,,
\\
\hat{\bm{e}}_{(\theta)} & = & \bm{[}0,0,\Sigma^{-1/2},0\bm{]}\,,
\\
\hat{\bm{e}}_{(\psi)} & = &
 \bm{[}0,0,0,{\cal A}^{-1/2}\Sigma^{1/2}\sin^{-1}{\!\theta}\bm{]}\,, 
\end{eqnarray}
where ${\cal K}(r)=\left[1+(2Mr)^{-1/2}(r^2+a^2)^{1/2}\right]^{-1}$.

\newpage
\References

\item
Abramowicz M A, Karas V, Lanza A 1998
``On the runaway instability of relativistic accretion tori'',
{\it{Astron.\ Astrophys.}} {\bf{}331} 1143--46

\item
Asseo E, Sol H 1987
``Extragalactic magnetic fields'',
{\it{Phys.\ Rep.}} {\bf{}148} 307--436

\item
Bi\v{c}\'ak J, Jani\v{s} V 1980
``Magnetic fluxes across black holes'',
{\it{Mon.\ Not.\ Roy.\ Astron.\ Soc.}} {\bf{}212} 899--915

\item
Blandford R D, Znajek R 1977
``Electromagnetic extraction of energy from Kerr black holes'',
{\it{Mon.\ Not.\ Roy.\ Astron.\ Soc.}} {\bf{}179} 433-56

\item
Bullard E C 1949
``Electromagnetic induction in a rotating sphere'',
{\it{Proc.\ Roy.\ Soc.\ Lond.}} {\bf{}199} 413--41

\item
Christodoulou D, Ruffini R 1973
``On the electrodynamics of collapsed objects'',
in {\it{}Black Holes}, eds C DeWitt \& B~S DeWitt
(New York: Gordon and Breach Science Publishers) p~R151

\item
Ciufolini I, Wheeler J A 1995
{\it{}Gravitation and Inertia} (New Jersey: Princeton University Press)

\item
Daigne F, Mochkovitch R 1997
``Gamma-ray bursts and the runaway instability of thick discs
around black holes'',
{\it{Mon.\ Not.\ Roy.\ Astron.\ Soc.}} {\bf{}285} L15--19

\item
Daigne F, Mochkovitch R 2000
``Mass loss from a magnetically driven wind emitted by a disk orbiting a
stellar mass black hole'',
in {\it{}Proceedings of the 5th Huntsville Gamma-Ray Burst Symposium}
(Huntsville 1999), in press; astro-ph/9912444

\item
Duncan R C, Thompson C 1992,
``Formation of very strongly magnetized neutron stars'',
{\it{Astrophys.\ J.\ Lett.}} {\bf{}392} L9--13

\item
Ghosh P, Abramowicz M A 1997
``Electromagnetic extraction of rotational energy from
disc-fed black holes'',
{\it{Mon.\ Not.\ Roy.\ Astron.\ Soc.}} {\bf{}292} 887--95

\item
Goldreich P, Julian W H 1969
``Pulsar electrodynamics'',
{\it{Astrophys.\ J.}} {\bf{}157} 869--80

\item
Hanami H 1997
``Magnetic cannonball model for gamma-ray bursts'',
{\it{Astrophys.\ J.}} {\bf{}491} 687--96

\item
Herzenberg A, Lowes F J 1957
``Electromagnetic induction in rotating conductors'',
{\it{Phil.\ Trans.\ Roy.\ Soc.\ Lond.}} {\bf{}249} 507--84

\item
Hewish A \etal\ 1968
``Observation of a rapidly pulsating radio source'',
{\it{Nature}} {\bf{}217} 709--13

\item
Jackson J D 1975 {\it{}Classical Electrodynamics} (New York: Wiley)

\item
King A R, Lasota J P, Kundt W 1975
``Black holes and magnetic fields'',
{\it{Phys.\ Rev.\ D}} {\bf{}12} 3037--42

\item
Kouveliotou C \etal\ 1998
``An X-ray pulsar with a superstrong magnetic field in the
soft gamma-ray repeater SGR 1806-20'',
{\it{Nature}} {\bf{}393} 235--37

\item
Kronberg P P 1994
``Extragalactic magnetic fields'',
{\it{Rep.\ Prog.\ Phys.}} {\bf{}57} 325

\item
Krotkov R V, Pellegrini G N, Ford N C, Swift A R 1999
``Relativity and the electric dipole moment of a
moving, conducting, magnetized sphere'',
{\it{Am.\ J.\ Phys.}} {\bf{}67} 493--98

\item
Lee H K, Wijers R A M J, Brown G E 2000
``The Blandford-Znajek process as a central engine for gamma-ray bursts'',
{\it{Phys.\ Rep.}} {\bf{}325} 83--114

\item
Mestel L 1999 {\it{Stellar Magnetism}} (Oxford: Clarendon Press)

\item
Michel F C, Hui Li 1999
``Electrodynamics of neutron stars'',
{\it{Phys.\ Rep.}} {\bf{}318} 227--97

\item
Mihara T \etal\ 1990
``New observations of the cyclotron absorption feature in
Hercules X-1'',
{\it{Nature}} {\bf{}346} 250--52

\item
Misner C W, Thorne K S, Wheeler J A 1973 {\it{}Gravitation}
(New York: W~H Freeman \& Co)

\item
Miyama S M, Tomisaka K, Hanawa T (eds) 1999
{\it{}Numerical Astrophysics} (Boston: Kluwer Academic Publishers)

\item
Murakami  T \etal\ 1988
``Evidence for cyclotron absorption from spectral features in
gamma-ray bursts seen with Ginga'',
{\it{Nature}} {\bf{}335} 234--35

\item
Pacini F 1967
``Energy emission from a neutron star'',
{\it{Nature}} {\bf{}216} 567--68

\item
Punsly B 1996
``Fast Waves and the Causality of Black Hole Dynamos''
{\it{Astrophys.\ J.}} {\bf{}467} 105--25

\item
Semer\'ak O 1993
{\it{Gen.\ Rel.\ Grav.}} {\bf{}25} 1041

\item
Thorne K S, Price R H, Macdonald D A 1986 {\it{Black Holes:
The Membrane Paradigm}} (New Haven: Yale University Press)

\item
Tomimatsu A 2000
``Relativistic dynamos in magnetospheres of rotating compact objects'',
{\it{Astrophys.\ J.}} {\bf{}528} 972--978; astro-ph/9908298

\item
Wald R M 1974
``Black hole in a uniform magnetic field'',
{\it{Phys.\ Rev.\ D}} {\bf{}10} 1680--85

\item
Zhang B, Harding A C 2000
``High magnetic field pulsars and magnetars: a unified picture'',
{\it{Astrophys.\ J.\ Lett.}} in press; astro-ph/0004067

\endrefs
\newpage

\newcommand{\ri}{{\rm{i}}}
\newcommand{\calE}{{\cal{E}}}

\title{Magnetic fluxes across black holes in a strong magnetic field regime}
\author{V.~Karas ~and 
 Z.~Bud\'{\iacc}nov\'a\\
 Astronomical Institute, Charles University Prague,\\
 V~Hole\v{s}ovi\v{c}k\'ach~2, CZ-180\,00~Praha, Czech~Republic}

\begin{abstract}
The magnetic flux across a magnetized Kerr-Newman black hole is
examined. It is shown by employing the Ernst-Wild spacetime how the flux
across an axially symmetric cap located on the horizon depends on the
magnetic field strength and vanishes for certain values of parameters,
analogously to the case of an external asymptotically uniform and
aligned test magnetic field across extreme black holes. Discussion in
terms of intuitive graphs is presented.

\end{abstract}

\date{PACS numbers: 97.60.Lf, 04.40.Nr}

\bigskip

\noindent
Journal: Physica Scripta {\bf 61}, 25 (2000)

\maketitle
\setcounter{equation}{0}
\setcounter{section}{0}
\renewcommand{\theequation}{\arabic{equation}}

\section{Introduction}
The structure of magnetic fields interacting with the gravitational
field of a black hole attracted considerable interest in the 1970s, and
since then the subject has reached a text-book level
\cite{TPM86}--\cite{G86}. It is of its own interest and helps to clarify
various issues within the framework of the classical Einstein-Maxwell
theory, namely the solution-generating techniques in general relativity
\cite{KSMH80,AG96}, but the main motivation to study black holes
immersed in magnetic fields originates in astrophysics. We know for sure
that interstellar and intergalactic magnetic fields do exist and they
get amplified in the course of accretion onto compact objects when the
field is frozen in plasma and the dynamo effects are involved
\cite{AS87,K94}. However, the energy density contained in realistic
magnetic fields, though they can be extremely strong (about $10^{12}$
gauss near neutron stars, and $\approx10^{15}$ gauss around presumed
magnetars \cite{HK98} and, possibly, in gamma-ray bursters \cite{L99}),
turns out to be too low to influence the background spacetime metric.
Test-field solutions are adequate in such circumstances, and
corresponding exact solutions of coupled Einstein-Maxwell equations are
mainly the question of principle. The influence of large-scale magnetic
fields upon the black-hole metric can be roughly characterized by
dimensionless parameter $\beta=B_0M$, where $B_0$ is the field strength
and $M$ is the mass of the object in geometrized units. Considering the
above-mentioned limit on the field strength near a one solar-mass
magnetar, we obtain $\beta\approx10^{-5}B_{15}(M/M_{\odot})\ll1$, where
$B_{15}=B_0/(10^{15}{\rm{gauss}})$. Such magnetic field contributes
substantially to the spacetime metric on spatial scales of
$r\approx\beta^{-1}$, as can be easily seen by direct inspection of
eq.~(\ref{g}) below. In physical units, this scale implies
$r\approx10^5$\,km.

Recently, the implications of the classical solutions have been resurrected 
by several authors within the framework of the string theory. For example,
superconducting properties of extremal solitonic objects ($p$-branes)
\cite{CEG98} can be related to the expulsion of external, asymptotically
uniform magnetic fields from extremal black holes (rotating and
electrically charged) in general relativity \cite{W74}--\cite{BJ85}. We
further discuss this effect for aligned magnetic fields in the 
next section. (There is apparently a
misinterpretation in Sec.\ V of \cite{CEG98} concerning the expulsion of
{\em{non-aligned\/}} test fields. These fields are not expelled even in
the limit of extreme black holes, as can be shown by calculating
magnetic fluxes \protect\cite{BJ85} and plotting the lines of magnetic
force \protect\cite{K89}.)

Here we rest on the grounds of classical exact solutions and we deal
with the stationary, aligned, asymptotically uniform magnetic fields
around magnetized Kerr-Newman (MKN) black holes \cite{EW76,D85}. Even
such simplified solutions are of interest: in a restricted region, they
can approximate collapsed objects interacting with the fields, which are
generated by remote sources. These calculations are also useful as
test-beds for checking numerical solvers. By employing the standard
covariant definition of the magnetic flux, otherwise complicated 
expressions can be illustrated in an intuitive manner. We will observe 
the expulsion of strong fields out of the horizon, as it occures
in the MKN spacetime with vanishing total angular momentum $J$
or vanishing total electric charge $Q$, both quantities being defined in 
terms of Komar's integrals \cite{GP78}--\cite{KV90}. One can ask whether 
the magnetic field is, 
in some natural way, separable into two parts, the first one
corresponding to a dipole-type field of the rotating body, and the other
one corresponding to the external field (generated by outside sources),
which gets expelled when the black hole becomes extreme.  In the present
paper we show that the aligned and uniform external magnetic field is
expelled out of the extreme MKN black hole in the limit of the weak
magnetic field, however, similar distinction of the field into the black 
hole's dipole-type part and the external part cannot be meaningfully 
introduced in the strong-field regime.

\section{The magnetic flux}

\subsection{Motivations}
Our present subject of the magnetized black holes can be linked to an
analogous problem in classical electrodynamics when a rotating sphere is
immersed in an external magnetic field. The reason for this analogy
stems from the fact that the black hole horizon can be treated as
a membrane with effective resistivity. The well-known similarity
between these two situations is instructive, and we thus briefly recall
several relevant works as the motivation for the present paper. The
classical problem was treated in original works by Faraday, Lamb,
Thomson and Hertz. More recently, Elsasser \cite{E46,E50} and Bullard
\cite{B48,B49} took interest in the theory of geomagnetic dynamo and
examined quasi-stationary solutions for a sphere or concentric spheres
with general orientation between rotation axis and the external field.
(Non-stationary oscillating electromagnetic fields are far more complicated,
and they decay rapidly with the relaxation time-scale which is of the
order of light-crossing time across the sphere; corresponding solutions
have been discussed also in context of black holes \protect\cite{MS85}.)
Geomagnetic applications enable one to ignore higher-order terms in
$v^2/c^2$ and to simplify the problem considerably. Herzenberg \& Lowes
\cite{HL57} extended previous works by detailed discussion of special
configurations, in particular, these authors examined the dragging of
asymptotically misaligned field-lines and their winding up onto a
rotating body. Schmutzer \cite{S77}, in a series of articles, discussed
asymptotically uniform stationary magnetic fields interacting with a
rotating and electrically conducting sphere, and he derived expressions
for the electric charge induced on the surface and for Foucalt currents
which, in turn, affect rotation of the sphere. Similar topics were applied
to the pulsar astrophysics by Pacini \cite{P67} and Goldreich \& Julian
\cite{GJ69}, and then developed by numerous authors till the present
\cite{M91}. Again, there is analogous winding effect which drags
non-aligned fields around black holes \cite{K89,PB77} and gives rise to
fictitious induced charges and currents on the horizon
\cite{RT73}--\cite{D78}. Astrophysically realistic pulsar magnetospheres
are more complicated, however, because in this case one cannot assume
the electrovacuum condition.

The special-relativity terms were examined by several authors. Bladel
\cite{B76,B84} performed standard but tedious expansions in small
parameter $v/c$, while Georgiou \cite{G82} formulated and solved this
problem within the approximation of perfect magnetohydrodynamics which
is particularly relevant in astrophysics. Numerical codes dealing with
general-relativistic magnetohydrodynamics have been developed and
applied to the problem of astrophysical accretion and jets
\cite{Y93,K98}. Finally, the structure of rotating neutron stars with
magnetic fields has been studied within full general relativity
\cite{BBGN95}. Very strong magnetic fields deform the shape of the star;
then, if the field is non-aligned with rotation axis and if the star
rotates rapidly, substantial amount of gravitational waves may be
produced. In the next section we restrict ourselves to much simpler
class of axisymmetric and stationary spacetimes.

\subsection{Magnetized black holes}
\label{mbh}
Magnetized Kerr-Newman metric \cite{EW76} represents an exact
electrovacuum solution. In spheroidal coordinates and in
the standard notation it has a form
\beq \!\!\!\!\!\!\!\!\!\!
g=|\Lambda|^2\Sigma\left(\Delta^{-1}\rd{r}^2+\rd{\theta}^2
-\Delta{A^{-1}}\rd{t}^2\right)+
|\Lambda|^{-2}\Sigma^{-1}A\sin^2\theta
\left(\rd{\phi}-\omega\rd{t}\right)^2,
\label{g}
\eeq
where $\Sigma=r^2+a^2\cos^2\theta$, $\Delta=r^2-2r+a^2+e^2$,
$A=(r^2+a^2)^2-{\Delta}a^2\sin^2\theta$ are functions from the
Kerr-Newman metric, while $\Lambda=1+\beta\Phi-\frac{1}{4}\beta^2\calE$
is given in terms of the Ernst complex potentials $\Phi(r,\theta)$ and
$\calE(r,\theta)$:
\begin{eqnarray}
\Sigma\Phi
 &=& ear\sin^2\theta-{\ri}e\left(r^2+a^2\right)\cos\theta, \\
\Sigma\calE
 &=& -A\sin^2\theta-e^2\left(a^2+r^2\cos^2\theta\right)
 \nonumber \\
 & & + 2{\ri}a\left[\Sigma\left(3-\cos^2\theta\right)+a^2\sin^4\theta-
 re^2\sin^2\theta\right]\cos\theta.
\end{eqnarray}
All quantities are expressed in geometrized units and multiplied by
suitable power of $M$, so that we deal with dimensionless variables.
The explicit form of the metric function $\omega(r,\theta)$ was given in
\cite{D85} but is not necessary for our discussion of magnetic fluxes.
The spacetime (\ref{g}) is characterized by parameters $a$, $e$, and
$\beta$ (angular-momentum parameter, electric-charge parameter, and the
magnetic field strength). The most general solution of this type
contains additional two parameters characterizing the asymptotic
electric field and magnetic charge of the black hole \cite{BD85}, but we
do not consider this possibility here.

The electromagnetic field can be written in terms of orthonormal components
in the locally non-rotating frame,
\begin{eqnarray}
H_{(r)}+{\ri}E_{(r)} &=& A^{-1/2}\sin^{-1}\!\theta\,\Phi^{\prime}_{,\theta},
\\
H_{(\theta)}+{\ri}E_{(\theta)} &=&
-\left(\Delta/A\right)^{1/2}\sin^{-1}\!\theta\,\Phi^{\prime}_{,r},
\label{heth}
\end{eqnarray}
where
$\Phi^{\prime}(r,\theta)
=\Lambda^{-1}\left(\Phi-\frac{1}{2}\beta\calE\right)$.
The horizon is located at $r{\equiv}r_+=1+\sqrt(1-a^2-e^2)$, independent of
$\beta$. As in the non-magnetized case, the horizon exists only for
$a^2+e^2\leq1$, equality corresponding to the extreme configuration.
Angular coordinates take the values $0\leq\theta\leq\pi$,
$0\leq\phi<2\pi|\Lambda_0|^2$ where
$|\Lambda_0|^2\equiv|\Lambda(\sin\theta=0)|^2=
1+\frac{3}{2}\beta^2e^2+2\beta^3ae+\beta^4\left(\frac{1}{16}e^4+a^2\right)$.

Knowing electromagnetic field structure one can calculate the total
electric charge $Q_{\rm{H}}$ and magnetic flux $F_{\rm{H}}(\theta)$
across a cap in axisymmetric position on the horizon (with the edge
defined by $\theta={\rm{const}}$):
\begin{eqnarray}
Q_{\rm{H}} &=& -|\Lambda_0|^2\,\Im{\rm{m}\,}\Phi^{\prime}\left(r_+,0\right),
\label{qh} \\
F_{\rm{H}} &=& 2\pi|\Lambda_0|^2\,\Re{\rm{e}\,}\Phi^{\prime}
 \left(r_+,\bar{\theta}\right)\Bigr|\strut^{\theta}_{\bar{\theta}=0}.
\label{fh}
\end{eqnarray}
These two quantities are defined invariantly and have been written in
different forms and approximations \cite{GP78}--\cite{KV90}. It is thus
useful to illustrate their properties by plotting graphs which in the
end appear very intuitive. Notice that $|\Lambda_0|^2$ is introduced in
eqs.\ (\ref{qh})--(\ref{fh}) by rescaling the range of azimuthal
coordinate, which is necessary to avoid a conical singularity on
symmetry axis \cite{H88,BK89}. (We allude to this fact because
the solution-generation technique guarantees that 
Einstein-Maxwell equations are satisfied by the spacetime
(\ref{g})--(\ref{heth}) locally; one has to further determine its 
global properties.) Given parameters $\beta$, $a$ and $e$,
the multiplicative factor $|\Lambda_0|$ is constant, independent of $r$.
It is thus convenient to rescale the variables containing this factor. We
introduce $Q=|\Lambda_0|^{-2}Q_{\rm{H}}$ and
$F=|\Lambda_0|^{-2}F_{\rm{H}}$ which do not contain the square of
$|\Lambda_0|$.

\begin{figure}[t]
 \epsfxsize=\hsize
 \centering
 \mbox{\epsfbox{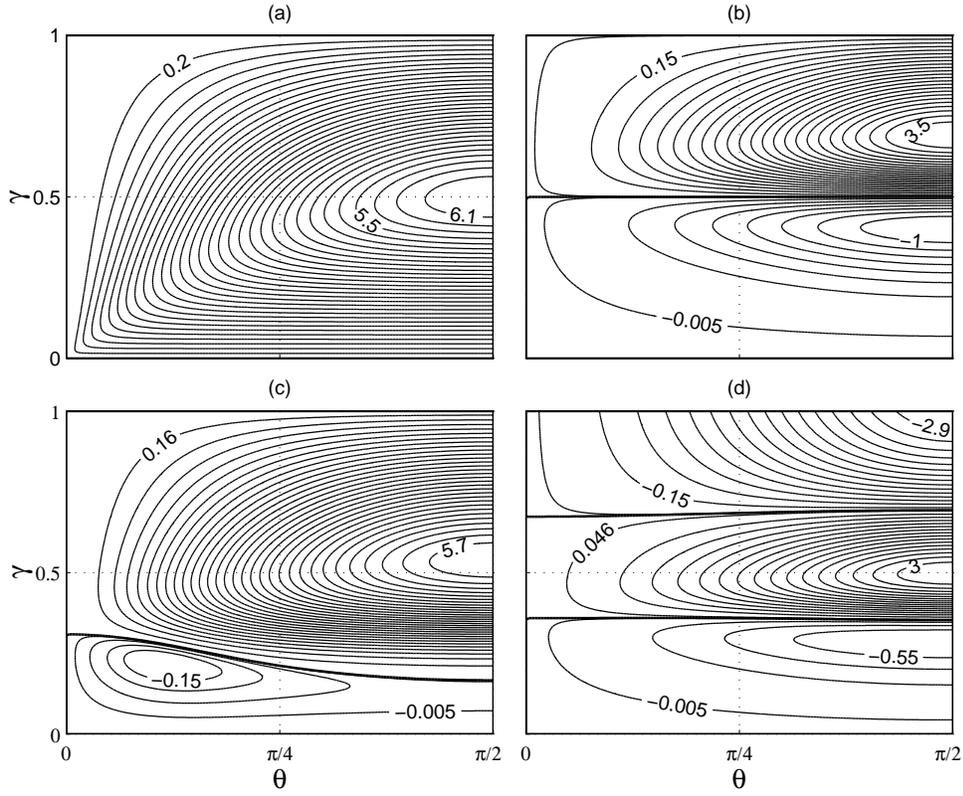}}
 \caption{The contour-lines of constant flux $F$ across a cap
 $\theta={\rm{const}}$ on the horizon for different values of
 magnetic-field parameter $\gamma(\beta)$. Parameters of the black hole
 are: (a)~$a=e=0$; (b)~$a=1$, $e=0$; (c)~$a=0.2$, $e=0$;
 (d)~$a=-e=1/\protect\sqrt{2}$. Selected values of $F$ are given with
 corresponding contours for orientation. See the text for details.
 \label{fig1}}
\end{figure}

\begin{figure}[t]
 \epsfxsize=\hsize
 \centering
 \mbox{\epsfbox{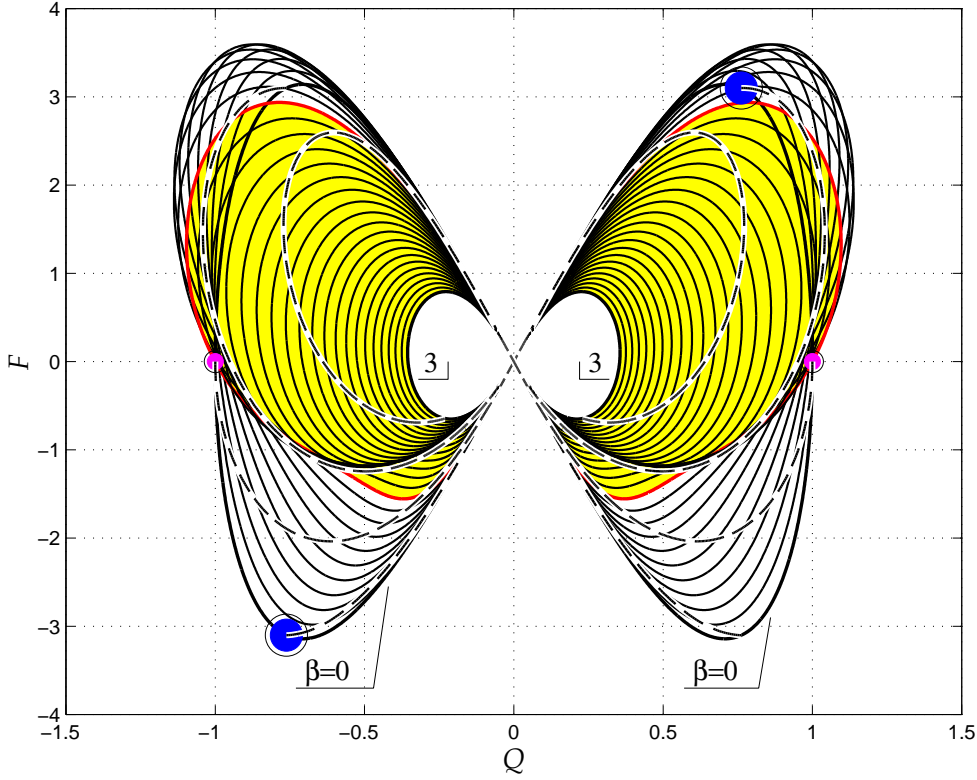}}
 \caption{The magnetic flux $F$ of the extreme magnetized Kerr-Newman black
 hole as a function of its electric charge $Q$. Each solid curve
 corresponds to a fixed value of $\beta$ within the range
 $\protect\langle0,3\protect\rangle$; $\beta=0$ is the case of
 Kerr-Newman black hole with no external magnetic field. The lines of
 constant ratio $a/e$ and varying $\beta$ are also plotted (dashed; the
 cases of $a/e=\pm0.85$ and $0$ are shown).
 \label{fig2}}
\end{figure}

\begin{figure}[t]
 \epsfxsize=\hsize
 \centering
 \mbox{\epsfbox{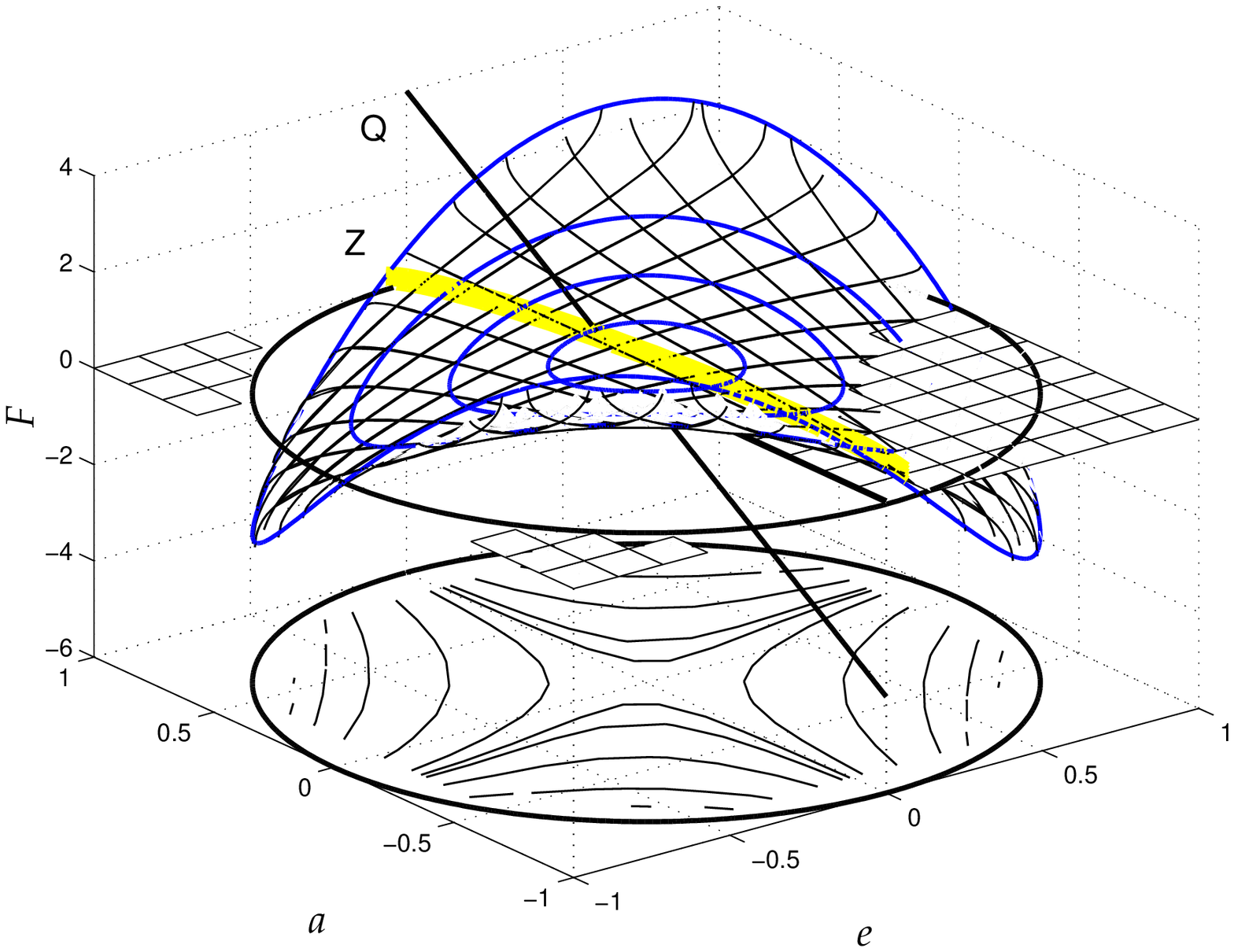}}
 \caption{Surface plot of the flux $F$ across the whole hemisphere
 $\theta=\pi/2$ as function of parameters $a$, $e$. The surface is
 defined on the circle $a^2+e^2\leq1$.
 \label{fig3}}
\end{figure}

Figure \ref{fig1} shows contours of constant $F$ in the plane of
$\gamma=\left(1+\beta\right)^{-1}$ and $\theta$ (size of the cap).
Panels (a)--(d) illustrate the topologically different situations which
arise for different combinations of $a$ and $e$. The whole range of
$0\leq\theta\leq\pi/2,$ $0\leq\beta(\gamma)\leq\infty$
($1\geq\gamma\geq0$) is captured in the four graphs. In particular,
setting $\theta=\pi/2$, one obtains the flux across the whole hemisphere.
Fig.\ \ref{fig1}a corresponds to the magnetized Schwarzschild metric
($a=e=0$) while (b) and (d) show two extreme configurations
($a^2+e^2=1$). One can see in Fig.\ \ref{fig1}a that the flux reaches
maximum of $F=2\pi$ for $\gamma=0.5$, the result which was mentioned
(apart from minor numerical errors) in refs.\ \cite{BJ85} and \cite{GP78}. 
In other words, the flux is {\em{not}\/} a monotonic function of the
magnetic-field parameter and it decreases when the field strength
exceeds certain value, depending on $a$ and $e$. (It is a monotonic 
function in the limit of a weak magnetic field only.) It turns out
that the flux is, in absolute value, less then $2\pi$ for all other
combinations of $a$, $e$. In terms of its $\beta$-dependence, the flux
first concentrates to symmetry axis $\theta=0$ when $\beta$ increases
from zero to unity, but than it spreads away from the axis. (It
was mentioned in ref.\ \protect\cite{BJ85} that this effect could
limit efficiency of the Blandford-Znajek mechanism for electromagnetic
extraction of energy from rotating black holes, because its power is
proportional to the flux. Characteristic value of
$\beta=1$ for which the flux is maximum, however, corresponds to
unrealistically strong magnetic fields for currently considered
astrophysical applications.) The extreme configurations (b) and (d) are more
complicated because the rotating charge of the black hole induces its own
contribution to the magnetic flux, and $F$ can even reverse its sign. The 
black hole's own contribution to the flux is visible particularly well in
Fig.\ \ref{fig1}d ($a=-e=1/\sqrt{2}$) where the contours intersect
abscissa $\gamma=1$. This happens in the Kerr-Newman case with $ae\neq0$
and no external magnetic field. The flux across {\em{any}\/} part of the
horizon vanishes for certain critical values of
$\beta=\beta_{\rm{cr}}(a,e)$, indicated by thick solid lines. For
example, $\beta_{\rm{cr}}(1,0)=1$ (cf.\ Fig.\ \ref{fig1}b). The exactly
horizontal direction of the separatrix corresponds to complete expulsion
of the field out of extreme black holes with vanishing charge $Q$ or
vanishing angular momentum $J$ \cite{D87,KV90}. What distinguishes 
the case (b) from (c) is extremality of the hole. The panel (c) 
corresponds to the rotating and charged (but not extreme) case;
again, the flux vanishes along the separatrix, but the magnetic field
still threads the horizon, $H_{(r)}(r=r_+)\neq0$, which in this graph
means that $\beta_{\rm{cr}}(a,e;\theta)$ depends on the cap's
latitudinal size rather than being constant.

Figure \ref{fig2} shows the curves of $F(Q)$ across the whole hemisphere
$\theta=\pi/2$ in the extreme MKN metric. Such graphs help us to elucidate
the behaviour of the flux when $\beta$ is kept fixed while the angular
momentum or the electric charge of the hole vary. Each of the
closed $\infty$-shaped curves corresponds to different strength of the
magnetic field (solid lines), and the part of the graph corresponding to
strong magnetic fields, $\beta\geq1$, is indicated by shading for
clarity. The two boundary curves are designated by their corresponding
values of $\beta$: ``$\beta=0$'' and ``$3$'' in the plot. Naturally, the
former one is anti-symmetric with respect to $Q=0$ because there is no
contribution to $F$ from external magnetic field. It is only the black
hole's rotating charge which gives rise to the magnetic flux when
$\beta=0$. In other words, each two points positioned on the ``0'' curve
anti-symmetrically with respect to origin are related by
$ae\rightarrow-ae$ in eq.\ (\ref{fh}). Two particular cases are
designated in the plot: (i)~$a=0$, $e=\pm1$ (small circles), and
(ii)~$a=0.65$, $e=\pm0.76$ (large circles). Notice that this feature of
anti-symmetrical $F(Q)$ is lost when the magnetic field becomes strong,
indicating that it would be difficult to define a natural split of the
total field near the horizon into a contribution of the rotating and charged
black hole and of the external field. Dashed curves correspond to
$a/e={\rm{const}}$. These curves start from a point on the
$\beta=0$ solid curve, they proceed through the origin of the plot (where
both $Q$ and $F(Q)$ vanish), and terminate at their intersection with the
$\beta=3$ curve. The intersection with the line $F=0$ at $Q\neq0$
corresponds to vanishing angular momentum, $J=0$. For example, the
dashed curve starting from the small circle corresponds to the magnetized
extreme Reissner-Nordstr{\o}m configuration. In spite of $a=0$, the flux
is in general non-zero along this curve in agreement with the fact of
non-vanishing $J$.

Figure \ref{fig3} is a surface plot of $F(a,e)_{|\theta=\pi/2}$ for a
fixed value of $\beta=0.05$ (projected contours are shown, too). The
surface is restricted by the condition $a^2+e^2\leq1$. Four circles of
$\sqrt(a^2+e^2)=0.25$, 0.5, 0.75, and 1.0 are shown. The shaded band on
the surface, denoted by ``{\sf{Z}}'', indicates where the total electric
charge is zero. Obviously, zero charge does not coincide with $e=0$,
and, on the other hand, $Q(a,e=0)$ does not vanish and its graph is
shown by solid curve ``{\sf{Q}}''. Here, $\beta$ is small and the curve
{\sf{Q}} looks almost linear (values are multiplied by factor of 40 for
clarity). As we have already mentioned, the flux across the extreme MKN 
hole vanishes also at other two points where $J$ is zero. These points are
near $a=0$ for $\beta\ll1$, as in this plot. On the other hand, absolute
value of $F$ is maximum for $|a|\approx|e|\approx1/\sqrt{2}$ because the
contribution to the flux from the black hole's rotating charge
dominates. The whole picture is qualitatively similar also for higher
values of $\beta$: the central part ($a^2+e^2\lta0.5$) of the surface,
however, is gradually pulled upwards (indicating that the contribution
of the hole's charge becomes less important), and the critical points
$Q=0$ and $J=0$ move away from $e=0$ and $a=0$ on the perimeter of the
unit circle.

The three above described figures contain complete information about
the electric charge of the hole and the magnetic flux threading its
horizon.

\section{Conclusions}
We examined the magnetic fluxes in the MKN spacetime for arbitrary
values of parameters $a$, $e$, and $\beta$. The form of the separatrix
in contour plots of $F$, as function of $\beta$ and of the polar-cap
size $\theta$, illustrates how the field is expelled out of the horizon,
and how a balance is established for certain
$\beta_{\rm{cr}}(a,e;\theta)$ between the external flux threading the
hole and the hole's own flux, which arises from its rotating charge.
Non-monotonic dependence of the flux on $\beta$ can be ascribed to the
interplay between two factors. First, the surface of the horizon
${\cal{S}}$ increases sharply with $\beta$ whenever $a$ and/or $e$ are
non-zero: ${\cal{S}}=4\pi\left(r^2_++a^2\right)|\Lambda_0|^2$. Second,
the magnetic field itself influences the flux distribution for high
values of $\beta$, as we illustrated by the graphs of magnetic fluxes
across the polar cap on the horizon. Let us recall that the term
$|\Lambda_0|^2$ has been introduced by rescaling the range of $\phi$ in
order to avoid the conical singularity on symmetry axis. Alternatively,
one may wish to accept the presence of this singularity and retain the
original azimuthal range of $2\pi$. In that case, the surface of the
horizon is independent of $\beta$. As a result of this rescaling, $|F|$
always starts decreasing for $\beta$ sufficiently large, while
$|F_{\rm{H}}|$ does not.

\medskip

{\protect\small
{\bf{Acknowledgements.}}
\def\lb#1{{\protect\linebreak[#1]}}
V\,K thanks for hospitality of the Group of Astronomy and
Astrophysics, G\"oteborg University and the Chalmers University of
Technology in Sweden, where this work was started. Support from the
grants GACR 205/\lb{2}97/\lb{2}1165, 202/\lb{2}99/\lb{2}0261, and GAUK
63/98 is acknowledged.}

\newpage

{\bf{References}}
\bigskip
                 

\end{document}